\documentclass[a4paper,11pt]{article}
\usepackage{jheppub}
\usepackage[latin1]{inputenc} 
\usepackage[T1]{fontenc}
\usepackage[safe]{textcomp}
\usepackage{lmodern} 
\usepackage{epsfig}
\usepackage{float}
\usepackage{amstext}  
\usepackage{amsfonts}
\usepackage{amsthm}
\usepackage{bm}       
\usepackage[bottom]{footmisc} 
\usepackage{array}                
\usepackage{xcolor} 
\usepackage{framed}
\usepackage{slashed}
\usepackage[absolute]{textpos}
\usepackage{axodraw4j}
\usepackage{dsfont}
\usepackage{multirow}

\newcommand{\p}{\partial}

\newcommand{\f}[2]{\frac{#1}{#2}}
\newcommand{\sss}[1]{\scriptscriptstyle#1}

\newcommand{\bea}{\begin{eqnarray}}
\newcommand{\eea}{\end{eqnarray}}
\newcommand{\be}{\begin{equation}}
\newcommand{\ee}{\end{equation}}
\newcommand{\ba}{\begin{align}}
\newcommand{\ea}{\end{align}}
\newcommand{\beas}{\begin{eqnarray*}}
\newcommand{\eeas}{\end{eqnarray*}}
\newcommand{\bes}{\begin{equation*}}
\newcommand{\ees}{\end{equation*}}
\newcommand{\bas}{\begin{align*}}
\newcommand{\eas}{\end{align*}}

\newcommand{\ssL}{{\mathcal L}} 
\newcommand{\eps}{{\varepsilon}}
\newcommand{\cd}{{\cdot}} 
\newcommand{\cf}{C_{\scriptscriptstyle{F}}} 
\newcommand{\ca}{C_{\scriptscriptstyle{A}}}
\newcommand{\tr}{T_{\scriptscriptstyle{F}}}

\newcommand{\dR}{d_{\scriptscriptstyle{R}}}
\newcommand{\Ng}{n_{\scriptscriptstyle{g}}}

\newcommand{\Nc}{N_{\scriptscriptstyle{c}}}
\newcommand{\Nf}{n_{\scriptscriptstyle{f}}}
\newcommand{\gs}{g_{\scriptscriptstyle{s}}}

\newcommand{\als}{\alpha_{\scriptscriptstyle{s}}}
\newcommand{\as}{a_{\scriptscriptstyle{s}}}
\newcommand{\lb}{\left(}
\newcommand{\rb}{\right)}

\definecolor{bluemar}{rgb}{0,0,.5}
\definecolor{redmar}{rgb}{.8,0,0}
\definecolor{greenmar}{rgb}{0,.5,0}

\def\bbuildrel#1_#2^#3%
{\mathrel{\mathop{\kern 0pt#1}\limits_{#2}^{#3}}}

\newcommand{\ice}[1]{\relax}
\newcommand{\beq}{\begin{equation}}
\newcommand{\eeq}{\end{equation}}
\newcommand{\re}[1]{(\ref{#1})}
\newcommand{\sbz}{  }

\newcommand{\ta}{\theta}

\newcommand{\rmi}[1]{{\mbox{\scriptsize #1}}}
\newcommand{\unit}{{\mathds{1}}} 

\title{OPE of the energy-momentum tensor correlator in massless QCD}
\author[a]{M. F. Zoller}
\author[a]{and K. G. Chetyrkin}
\affiliation[a]{Institut f\"ur Theoretische Teilchenphysik, Karlsruhe
  Institute of Technology (KIT), \mbox{D-76128 Karlsruhe, Germany}}

\emailAdd{max.zoller@kit.edu}
\emailAdd{Konstantin.Chetyrkin@kit.edu}

\abstract{We analytically calculate higher order corrections to coefficient
functions of the operator product expansion (OPE) for the Euclidean
correlator of two energy-momentum tensors in massless QCD.
These are the three-loop contribution to the coefficient $C_0$ in front of the unity operator
$O_0=\mathds{1}$ and the one and two-loop contributions  to the coefficient $C_1$
in front of the gluon ``condensate'' operator $O_1=-\f{1}{4} G^{\mu \nu}G_{\mu \nu}$.
For the correlator of two operators $O_1$ we  present the coefficient $C_1$ at two-loop level
(the coefficient function  $C_0$ is known at four loops from 
\cite{Baikov:2006ch}).
}

\keywords{QCD, Quark-Gluon Plasma, Sum Rules}
\arxivnumber{1209.1516}
\subheader{TTP12-025\\SFB/CPP-12-56}

\begin{document}
\maketitle

\setlength{\fboxrule}{0.5 mm} 

\section{Motivation}
The energy-momentum tensor correlator 
\be 
T^{\mu\nu;\rho\sigma}(q)= i\int\!\mathrm{d}^4x\,e^{iqx}\, \langle 0 | \hat{T}^{\mu\nu;\rho\sigma}(x)| 0\rangle,
\ \
 \hat{T}^{\mu\nu;\rho\sigma}(x) = T[ T^{\mu\nu}(x)T^{\rho\sigma}(0)]
\label{TT}
\ee
plays an important role in many physical problems.
A lot of these lie in the field of Quark Gluon Plasma (QGP) physics.
Here the correlator eq.~\re{TT} is the central object for describing transport properties, like the shear viscosity of the plasma
(see e.g. \cite{Meyer:2008sn,Meyer:2007ic}) and spectral functions for some tensor channels in the QGP \cite{Meyer:2008gt}.
Another application is a sum rule approach to tensor glueballs. These special hadrons without valence quarks are determined
by their gluonic degrees of freedom. QCD allows for such particles but a conclusive discovery has not yet been made. 

In a sum rule approach \cite{Shifman:1978bx} one usually starts with
the vacuum correlator of an interpolating local operator which has the
same quantum numbers as the hadrons we want to investigate. If we are
interested in glueballs we take local operators consisting of gluon
fields. For the cases $J^{PC}=0^{++},0^{-+}$ and $2^{++}$ the
following operators are usually considered:
 \begin{align}
     O_1(x)  &=-\f{1}{4} G^{\mu \nu}G_{\mu \nu}(x)  & \text{(scalar)} \label{O1}\\
     \tilde{O}_1(x)  &=G^{\mu \nu}\tilde{G}_{\mu \nu}(x)  & \text{(pseudoscalar)}\label{O1t}\\ 
     O_T(x) &=T^{\mu \nu}(x) & \text{(tensor)}
    \end{align} 
where $G_{\mu \nu}$ is the gluon field strength tensor and
\be
\tilde{G}_{\mu \nu}=\eps_{\mu\nu\rho\sigma}G^{\rho \sigma}
\ee
the dual gluon field strength tensor. For more details see
\cite{forkel_sumrule}. The vacuum expectation value (VEV) of the
correlator of such a local operator $O(x)$
\be 
\Pi(Q^2)=i\int\!\mathrm{d}^4x\,e^{iqx}\,\langle 0|T[\,O(x)O(0)]|0 \rangle \qquad (Q^2=-q^2) 
\label{correlator}
\ee
can of course be calculated in perturbation theory for large Euclidean momenta, but this is not enough.
Starting from the perturbative region of momentum space we can probe into the non-perturbative region by means
of an OPE. The idea originally formulated in
\cite{wilson_ope} is to expand the non-local operator product $i\int\!\mathrm{d}^4x\,e^{iqx}\,T\,[O(x)O(0)]$
in a series of local operators with Wilson coefficients depending on the large Euclidean momentum q. In sum rules
we usually have dispersion relations connecting the VEV of such a Euclidean operator product to some spectral density
in the physical region of momentum space.
As we are ultimately interested in the VEV of this operator product we only have to consider gauge invariant 
scalar operators in the expansion.
\ice{
\footnote{The vacuum is defined as a 
gauge and Lorentz invariant as well as colourless state.} 
}

Effectively this expansion separates the high energy physics, which is
contained in the Wilson coefficients, from the low energy physics which
is taken into account by the VEVs of the local operators, the
so-called condensates \cite{Shifman:1978bx}. These cannot be calculated in perturbation
theory, but need to be derived from low energy theorems or be
calculated on the lattice.
\ice{\footnote{In fact, in perturbation theory
applied to massless QCD the VEV of operators like $O_1$ vanishes,
whereas it has been argued that in the low energy range there is a
finite value for this so-called gluon condensate, see
e.g. \cite{Novikov_scalargluonium}.}
}
Such an OPE has already been
done for the cases eq.~\re{O1} and eq.~\re{O1t} (see
\cite{Novikov_scalargluonium, Novikov1979347}) with one-loop accuracy.

In this work we  present the results for the Wilson coefficients in front of the
operators $O_0$ and $[O_1]$ for the correlator eq.~\re{TT} in massless QCD:
\be \hat{T}^{\mu\nu;\rho\sigma}(q)
\bbuildrel{=\!=\!=}_{q^2 \to -\infty}^{} 
C_{0}^{\mu\nu;\rho\sigma}(q) \mathds{1} + C_{1}^{\mu\nu;\rho\sigma}(q) [O_1]+\ldots .
\label{TTexp1}\ee
\vskip 2ex 
The brackets in $[O_1]$ indicate that we take a renormalized form of the operator $O_1$:
\be [O_1]=Z_G O_1^B=-\f{Z_G}{4}G^{B\,\mu \nu}G^{B}_{\mu \nu}\ee
where the index $B$ marks bare quantities.
We start our calculation with bare quantities which are expressed through renormalized ones in the end:
\bea 
 \hat{T}^{\mu\nu;\rho\sigma}(q)&=&\sum \limits_i C_{i}^{B\,\mu\nu;\rho\sigma}(q) O_{i}^B
\label{TTexp2_B}
\\
&=&\sum \limits_i C_{i}^{\mu\nu;\rho\sigma}(q) [O_{i}].
\label{TTexp2_R}
\eea
All physical matrix elements of $[O_1]$ are finite and so is the renormalized 
coefficient\footnote{This statement  as well as eqs.~\re{TTexp2_B} and eqs.~\re{TTexp2_R}
are only true modulo so-called {\em contact terms}; see a detailed discussion
in the next section.}
\be C_1=\f{1}{Z_G}C_1^B. \label{C1ren}
\ee
The renormalization constant
\be Z_G=1+\als\f{\p}{\p\als}\ln Z_{\als}=\lb 1- \f{\beta(\als)}{\eps}\rb^{-1} \ee
has been derived in a simple way in  \cite{Spiridonov:1984br} (see also an earlier work
\cite{Nielsen:1975ph}).
Here $Z_{\als}$ is the
renormalization constant for $\als$ and we define\footnote{Often in the
literature $Z_{\als}$ is used instead of $Z_G$ and $\als G^{\mu
\nu}G_{\mu \nu}$ instead of $O_1$. This is justified because up to
first order in $\als$ the renormalization constants $Z_G$ and
$Z_{\als}$ are the same.  Only in higher orders $Z_G$ and $Z_{\als}$
differ and therefore $Z_G$ has to be used in such cases.}
\beq\beta(\als)=
\mu^2\f{\mathrm{d}}{\mathrm{d}\mu^2}\, \ln \als
=  - \sum_{i \ge 0} \beta_i \, \left( \frac{\als}{\pi} \right)^{i+1}
{}.
\label{be:def}
\eeq

In the massive case the two and four dimensional operators $O_f=m_f^2$
and $O_2^f=m_f\bar{\psi_f}\psi_f$ would have to be included as well
for every massive quark flavour f. The VEVs of all other linearly
independent scalar operators of dimension four vanish either by some
equation of motion or as they are not gauge invariant. The
contributions of higher dimensional operators are suppressed by higher
powers of $\f{1}{Q^2}$ in the coefficients.

\ice{\footnote{Applying an
appropriate weighting, like e.g. the Borel transformation, during the
sum rule calculation the suppression of higher dimensional operators
becomes even more efficient.}}

Apart from the leading coefficient
$C_0$ in front of the local operator $O_0=\mathds{1}$ the coefficient
$C_1$ in front of $[O_1]$ is of special interest for many
applications.  One example is if we have a spectral density defined by
our correlator and we want to calculate the shift in this spectral
density from zero to finite temperature:
\be \Delta \rho (\omega,T)=\rho(\omega,T)-\rho (\omega,0). \label{deltarho}\ee
The spectral density at $T=0$ is calculated from the VEV of the
correlator whereas for the spectral density at finite T we take the
thermal average of the operator product. For the unity operator
$O_0=\mathds{1}$ the VEV and the thermal average are both $1$ due to
the normalization conditions. Hence the leading term from the OPE,
i.e. the one proportional to $O_0$ vanishes in eq.~\re{deltarho} which
makes the Wilson coefficients in front of $O_1$ and $O_2^f$ the
leading high frequency contributions to eq.~\re{deltarho}. For more
details see e.g. \cite{finite_temp_OPE}.

\section{The energy-momentum tensor in QCD}
The energy-momentum tensor which can be derived from the Lagrangian of
a field theory is an interesting object by  itself.  To be identified
with the physical object known from classical physics and general
relativity it has to be symmetric as well as conserved. A very general
method to derive such an energy-momentum tensor can be found e.g. in
\cite{Freedman197495,Freedman1974354,weinberg_cosmo}. This has firstly
been done for QCD in \cite{nielsen_Tmunu} and the result derived from
the renormalized Lagrangian
\be
\begin{split}
\ssL=&-\f{1}{4}Z_3\,G_{\mu \nu} G^{\mu \nu}-\f{1}{2\lambda}\lb\p_\mu A^{\mu}\rb^2 
+\tilde{Z_3}\p_\rho \bar{c} \p^{\rho}c+\gs\tilde{Z_1}\p_\rho \bar{c}\lb A^{\rho} \times c\rb\\
&+\f{i}{2}Z_2 \bar{\psi}\overleftrightarrow{\slashed{\p}}\psi + \gs Z_{1\psi}\bar{\psi}\slashed{A} T\psi
\end{split} 
\label{2Lbare} 
\ee
is
\be
\begin{split}
T_{\mu \nu}=&-Z_3 G_{\mu \rho}G_{\nu}^{\,\,\rho}
+\f{1}{\lambda}(\p_\mu \p_\rho A^\rho)A_\nu+\f{1}{\lambda}(\p_\nu \p_\rho A^\rho)A_\mu \\
&+\tilde{Z_3}(\p_\mu \bar{c} \p_\nu c +\p_\nu \bar{c} \p_\mu c)
+g\tilde{Z_1}\left(\p_\mu \bar{c}(A_\nu \times c)+\p_\nu \bar{c}(A_\mu \times c)\right) \\
&+\f{i}{4}Z_2 \bar{\psi}\left(\overleftrightarrow{\p_\mu} \gamma_\nu+\overleftrightarrow{\p_\nu} \gamma_\mu\right)\psi
+\f{g}{2}Z_{1\psi}\bar{\psi}\left(A_\mu T \gamma_\nu+A_\nu\ T \gamma_\mu\right)\psi \\
&-g_{\mu\nu}\left\{ -\f{1}{4}Z_3\,G_{\rho \sigma} G^{\rho \sigma}
+\f{1}{\lambda}\lb\p_\sigma \p_\rho A^{\rho}\rb A^\sigma + \f{1}{2\lambda}\lb\p_\rho A^{\rho}\rb^2 \right.\\
&+\left.\tilde{Z_3}\p_\rho \bar{c} \p^{\rho}c+g\tilde{Z_1} \left(\p_\rho \bar{c}(A_\rho \times c)\right)
+\f{i}{2}Z_2 \bar{\psi}\overleftrightarrow{\slashed{\p}}\psi + gZ_{1\psi}\bar{\psi}\slashed{A}T\psi \right\},
\end{split}
\label{5Tqcd}
\ee
Here 
\be
G_{\mu \nu}=
\p_\mu A_\nu - \p_\nu A_\mu + \f{\tilde{Z_1}}{\tilde{Z_3}}\gs \lb A_\mu \times A_\nu\rb
{},
\ee
 $Z_3, \tilde{Z_3}$ and $Z_2$ stand for the field  renormalization constants for the gluon, ghost and quark fields
respectively and $\tilde{Z_1}$ and $Z_{1\psi}$ for  the vertex renormalization constants. 
The abbreviation
$\lb A_\mu \times A_\nu\rb^a=f^{abc}A^b_\mu A^c_\nu$, where $f^{abc}$ is the structure constant of the SU($\Nc$) gauge group,
is used and all colour indices are suppressed for convenience.\\
This energy-momentum tensor consists of gauge invariant as well as gauge and ghost terms. 
If we were to consider general matrix elements of
operator products we would have to include all these terms.
It has been pointed out in \cite{nielsen_Tmunu} however
that for Green's functions with only gauge invariant operators it would be enough to take the gauge invariant part of the energy
momentum tensor: 
\be
\begin{split}
T_{\mu \nu}|_{\sss{\mathrm{ginv}}}=&-Z_3 G_{\mu \rho}G_{\nu}^{\,\,\rho}
+\f{i}{4}Z_2 \bar{\psi}\left(\overleftrightarrow{\p_\mu} \gamma_\nu+\overleftrightarrow{\p_\nu} \gamma_\mu\right)\psi
+\f{g}{2}Z_{1\psi}\bar{\psi}\left(A_\mu T \gamma_\nu+A_\nu\ T \gamma_\mu\right)\psi \\
&-g_{\mu\nu}\left\{ -\f{1}{4}Z_3\,G_{\rho \sigma} G^{\rho \sigma}
+\f{i}{2}Z_2 \bar{\psi}\overleftrightarrow{\slashed{\p}}\psi + gZ_{1\psi}\bar{\psi}\slashed{A}T\psi \right\}.
\end{split}
\label{5Tqcdginv}
\ee
This has been checked in our calculation of $C_0$ which we have done once with the full energy-momentum tensor eq.~\re{5Tqcd} and once with
the gauge invariant part eq.~\re{5Tqcdginv} up to three-loop accuracy. As expected both calculations yield the same result.\\
The insertion of a local operator into a Green's function corresponds to an additional vertex in every possible Feynman diagram.
For the energy-momentum tensor eq.~\re{5Tqcd} we get the vertices  shown in Figure \re{fig_feynmanrules} \\[1ex]
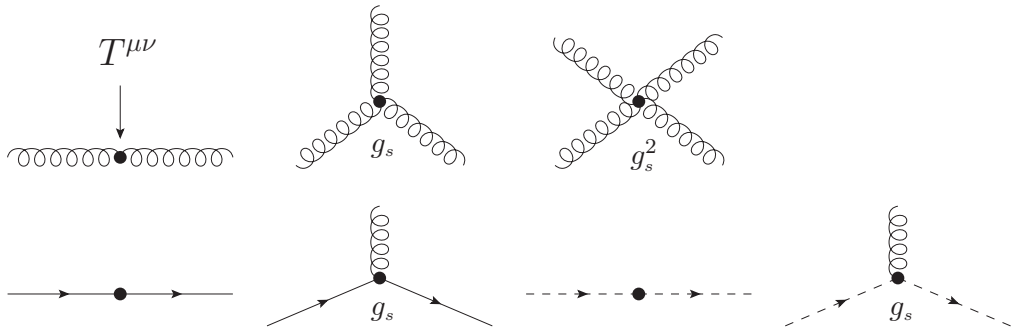
\begin{figure}[h!]
\begin{tabular}{cccc}
  \begin{picture}(85,30) (0,0)
\SetScale{0.6}
    \SetWidth{0.5}
    \SetColor{Black}
    \Gluon(10,25)(80,25){5.5}{6.5}
    \Gluon(80,25)(150,25){5.5}{6.5}
    \Vertex(80,25){4}
    \LongArrow(80,70)(80,40)
    \Text(40,50)[lb]{\Large{\Black{$T^{\mu \nu}$}}}
  \end{picture}
&
  \begin{picture}(85,70) (0,0)
\SetScale{0.6}
    \SetWidth{0.5}
    \SetColor{Black}
    \Gluon(80,60)(80,120){5.5}{6.5}
    \Gluon(80,60)(132,20){5.5}{6.5}
    \Gluon(80,60)(28,20){5.5}{6.5}
    \Vertex(80,60){4}
    \Text(45,15)[lb]{\Black{$\gs$}}
  \end{picture}
&
  \begin{picture}(85,70) (0,0)
\SetScale{0.6}
    \SetWidth{0.5}
    \SetColor{Black}
    \Gluon(80,60)(28,100){5.5}{6.5}
    \Gluon(80,60)(132,100){5.5}{6.5}
    \Gluon(80,60)(132,20){5.5}{6.5}
    \Gluon(80,60)(28,20){5.5}{6.5}
    \Vertex(80,60){4}
    \Text(45,10)[lb]{\Black{$\gs^2$}}
  \end{picture}
 & \\
  \begin{picture}(85,30) (0,0)
\SetScale{0.6}
    \SetWidth{0.5}
    \SetColor{Black}
    \ArrowLine(10,25)(80,25)
    \ArrowLine(80,25)(150,25)
    \Vertex(80,25){4}
  \end{picture}
&
  \begin{picture}(85,47) (0,0)
\SetScale{0.6}
    \SetWidth{0.5}
    \SetColor{Black}
    \ArrowLine(10,5)(80,35)
    \ArrowLine(80,35)(150,5)
    \Gluon(80,35)(80,80){5.5}{4.5}
    \Vertex(80,35){4}
    \Text(45,5)[lb]{\Black{$\gs$}}
  \end{picture}
&  
  \begin{picture}(85,35) (0,0)
\SetScale{0.6}
    \SetWidth{0.5}
    \SetColor{Black}
    \DashArrowLine(10,25)(80,25){5}
    \DashArrowLine(80,25)(150,25){5}
    \Vertex(80,25){4}
  \end{picture}
&
  \begin{picture}(85,47) (0,0)
\SetScale{0.6}
    \SetWidth{0.5}
    \SetColor{Black}
    \DashArrowLine(10,5)(80,35){5}
    \DashArrowLine(80,35)(150,5){5}
    \Gluon(80,35)(80,80){5.5}{4.5}
    \Vertex(80,35){4}
    \Text(45,5)[lb]{\Black{$\gs$}}
  \end{picture}
 \\
\end{tabular}
\caption{Energy-momentum tensor vertices and their dependence on $\gs$}
\label{fig_feynmanrules}
\end{figure} 

In \cite{nielsen_Tmunu} it has been proven that the energy-momentum
tensor of QCD is a finite operator which means the Z-factors appearing in  \re{5Tqcd}
make any Green function of (renormalized) QCD elementary fields with {\em one} insertion of the operator
$T_{\mu \nu}$ finite.
\ice{ or the bilocal operator   $\hat{T}^{\mu\nu;\rho\sigma}(x)$. } We have used this theorem as a  check for our setup and have
calculated one - and two-loop corrections to the matrix elements $\langle \text{gluon(p,$\mu_1$)}
|\hat{T}^{\mu\nu}|\text{gluon(p,$\mu_2$)}\rangle$, $\langle \text{ghost(p)}
|\hat{T}^{\mu\nu}|\text{ghost(p)}\rangle$ and $\langle \text{quark(p),
gluon(0,$\mu_1$)} |\hat{T}^{\mu}_{\,\,\mu}|\text{quark(p)}\rangle$ which
turned out to be finite as expected.

Another important consequence of the finiteness property is the
absence of the anomalous dimension of the energy-momentum
tensor. For the bilocal operator $\hat{T}^{\mu\nu;\rho\sigma}(x)$ the
situation is more complicated. This is because of extra 
(quartic!) UV divergences appearing  in the limit of $x\to 0$.
If $x$ is kept away from 0 then $\hat{T}^{\mu\nu;\rho\sigma}(x)$ is finite and renormalization scheme independent.  
These divergences (which are local in $x$!) manifest themselves in the Fourier transform  $T^{\mu\nu;\rho\sigma}(q)$.
They can and should be renormalized with proper counterterms:
\beq
[T^{\mu\nu;\rho\sigma}(q)] = T^{\mu\nu;\rho\sigma}(q) - \sum_i Z^{ct}_i(q) O_i
{},
\label{renorm:extra}
\eeq
where  $O_i$ are some operators of (mass) dimension $\le 4$ and $ Z^{ct}_i(q)$ are the corresponding
(divergent)  Z-factors.  The latter must be {\em local}, that is have  only {\em  polynomial }
dependence of the external momentum $q$. Within the $\overline{\text{MS}}$-scheme  $ Z^{ct}_i(q)$ are just poles in $\eps$.
It is of importance to note that the subtractive renormalization encoded in eq.~\re{renorm:extra} 
is in general {\em not} constrained by the QCD charge renormalization. Thus, 
the unambiguous QCD predictions for the coefficient functions in OPE \re{TTexp1} could be made only
modulo contact terms proportional to $\delta(x)$ in position space.

\ice{
Such  kind of divergences are  known
from the birth of Quantum Field Theory. 

For instance, in the correlator of two vector current $T[j_{\mu}(x) j{\nu}(0)]$ there exists a quadratic
UV divergence which, due to  the Ward  identity, is  proprtiin 
}

\section{OPE of the energy-momentum tensor correlator}
The leading coefficient $C_0$ is just the perturbative VEV of the correlator eq.~\re{TT} 
\be
C_0^{\mu\nu;\rho\sigma}(q)=\left.\langle 0| \hat{T}^{\mu\nu;\rho\sigma}(q)|0\rangle\right|_{\text{pert}}
\label{C0tensor}
\ee
which we have computed up to order $\als^2$ (three loops).  In Figure
\re{O0_dias} we show some sample Feynman diagrams contributing to
this calculation. The energy-momentum tensor plays the role of an
external current. In order to produce all possible Feynman diagrams we
have used the program QGRAF \cite{QGRAF}.  As these diagrams are
propagator-like the relevant integrals can be computed with the FORM
package MINCER \cite{MINCER} after projecting them to scalar
pieces. For the colour part of the diagrams the FORM package COLOR
\cite{COLOR} has been used.
\begin{figure}[h!]
\begin{center}
$
\langle 0|  \hat{T}^{\mu\nu;\rho\sigma}(q)|0 \rangle_{pert}=
  \begin{picture}(80,30) (0,0)
    \SetWidth{0.5}
    \SetColor{Black}
\Photon(5,0)(20,0){3}{2.5}
\Photon(60,0)(75,0){3}{2.5}
\LongArrow(7,8)(15,8)
\LongArrow(65,8)(73,8)
\CCirc(40,0){20}{Black}{Blue}
    \Text(5,-15)[lb]{{\Black{$\mu \nu$}}}
    \Text(60,-15)[lb]{{\Black{$\rho \sigma$}}}
    \Text(7,12)[lb]{{\Black{$q$}}}
    \Text(64,12)[lb]{{\Black{$q$}}}
  \end{picture}$\\[4ex]
$=
  \begin{picture}(80,30) (0,0)
    \SetWidth{0.5}
    \SetColor{Black}
\Photon(5,0)(20,0){3}{2.5}
\Photon(60,0)(75,0){3}{2.5}
\GlueArc(40,0)(20,0,180){3}{9.5}
\GlueArc(40,0)(20,180,360){3}{9.5}
    \Vertex(20,0){3}
    \Vertex(60,0){3}
  \end{picture}
+
  \begin{picture}(80,30) (0,0)
    \SetWidth{0.5}
    \SetColor{Black}
\Photon(5,0)(20,0){3}{2.5}
\Photon(60,0)(75,0){3}{2.5}
\ArrowArc(40,0)(20,0,180)
\ArrowArc(40,0)(20,180,360)
    \Vertex(20,0){3}
    \Vertex(60,0){3}
  \end{picture}
+
  \begin{picture}(80,30) (0,0)
    \SetWidth{0.5}
    \SetColor{Black}
\Photon(5,0)(20,0){3}{2.5}
\Photon(60,0)(75,0){3}{2.5}
\DashArrowArc(40,0)(20,0,180){5}
\DashArrowArc(40,0)(20,180,360){5}
    \Vertex(20,0){3}
    \Vertex(60,0){3}
   \end{picture}$\\[6ex]
$+
\begin{picture}(80,30) (0,0)
    \SetWidth{0.5}
    \SetColor{Black}
\Photon(5,0)(20,0){3}{2.5}
\Photon(60,0)(75,0){3}{2.5}
\GlueArc(40,0)(20,0,90){3}{4.5}
\GlueArc(40,0)(20,90,180){3}{4.5}
\GlueArc(40,0)(20,180,270){3}{4.5}
\GlueArc(40,0)(20,270,360){3}{4.5}
\Gluon(40,20)(40,-20){3}{6.5}
    \Vertex(40,20){2}
    \Vertex(40,-20){2}
    \Vertex(20,0){3}
    \Vertex(60,0){3}
\end{picture}
+
  \begin{picture}(80,30) (0,0)
    \SetWidth{0.5}
    \SetColor{Black}
\Photon(5,0)(20,0){3}{2.5}
\Photon(60,0)(75,0){3}{2.5}
\ArrowArc(40,0)(20,0,90)
\ArrowArc(40,0)(20,90,180)
\ArrowArc(40,0)(20,180,270)
\ArrowArc(40,0)(20,270,360)
\Gluon(40,20)(40,-20){3}{6.5}
    \Vertex(40,20){2}
    \Vertex(40,-20){2}
    \Vertex(20,0){3}
    \Vertex(60,0){3}
  \end{picture}
+
  \begin{picture}(80,30) (0,0)
    \SetWidth{0.5}
    \SetColor{Black}
\Photon(5,0)(20,0){3}{2.5}
\Photon(60,0)(75,0){3}{2.5}
\GlueArc(40,0)(20,0,90){3}{4.5}
\GlueArc(40,0)(20,90,180){3}{4.5}
\GlueArc(40,0)(20,180,270){3}{4.5}
\GlueArc(40,0)(20,270,360){3}{4.5}
\Gluon(40,20)(40,10){3}{1}
\Gluon(40,-10)(40,-20){3}{1}
    \Vertex(40,20){2}
    \Vertex(40,-20){2}
    \Vertex(40,10){2}
    \Vertex(40,-10){2}
\ArrowArc(40,0)(10,90,270)
\ArrowArc(40,0)(10,270,90)
    \Vertex(20,0){3}
    \Vertex(60,0){3}
  \end{picture}$\\[7ex]$+\ldots$
\caption{Diagrams for the calculation of the coefficient $C_0$} \label{O0_dias}
\end{center}
\end{figure}
Because of the four independent external Lorentz indices there are many possible tensor structures for the correlator eq.~\re{TT} 
and hence the Wilson coefficients. These are composed of the large external momentum q and the metric tensor g.
Using the symmetries\footnote{
These symmetries are  $\mu \longleftrightarrow \nu$,$\rho \longleftrightarrow \sigma$
and $(\mu \nu) \longleftrightarrow (\rho \sigma)$.}
of eq.~\re{TT} we can narrow them down to five possible independent tensor structures:
\be \begin{split}
t_1^{\mu\nu;\rho\sigma}(q)&=q^\mu q^\nu q^\rho q^\sigma ,\\
t_2^{\mu\nu;\rho\sigma}(q)&=q^2 \ \lb q^{\mu} q^{\nu} g^{\rho \sigma}+q^{\rho} q^{\sigma} g^{\mu \nu} \rb ,\\
t_3^{\mu\nu;\rho\sigma}(q)&=q^2 \ \lb q^{\mu} q^{\rho} g^{\nu \sigma}+q^{\mu} q^{\sigma} g^{\nu \rho} 
  +q^{\nu} q^{\rho} g^{\mu \sigma}+q^{\nu} q^{\sigma} g^{\mu \rho} \rb ,\\
t_4^{\mu\nu;\rho\sigma}(q)&=\lb q^2 \rb^2g^{\mu\nu}g^{\rho\sigma} ,\\
t_5^{\mu\nu;\rho\sigma}(q)&=\lb q^2 \rb^2 \lb g^{\mu \rho} g^{\nu \sigma}+g^{\mu \sigma} g^{\nu \rho}\rb.     
    \end{split} 
\label{t1_t5}
\ee
The conservation of the energy-momentum tensor leads to additional restrictions:
\be 
q_\mu\,T^{\mu\nu;\rho\sigma}(q)= \text{(local) contact terms} 
\label{Tmumu:conservation}
\ee
This condition has been checked in all calculations. Subtracting the physically irrelevant contact terms
leads to only two independent tensor structures which have also been used e.g. in \cite{Pivovarov_tensorcurrents}:
\be \begin{split}
                        t_S^{\mu\nu;\rho\sigma}(q)=&\eta^{\mu\nu} \eta^{\rho\sigma} \\
t_T^{\mu\nu;\rho\sigma}(q)=&\eta^{\mu\rho} \eta^{\nu\sigma}
                      +\eta^{\mu\sigma} \eta^{\nu\rho}
                      -\f{2}{D-1}\eta^{\mu\nu} \eta^{\rho\sigma} \\[2ex]
\text{with} \quad 
\eta^{\mu\nu}(q)=&q^2 g^{\mu\nu} - q^\mu q^\nu
\label{tST}
                       \end{split} \ee
where $D$ is the dimension of the space time.
The structure $t_T^{\mu\nu;\rho\sigma}(q)$ is traceless and orthogonal to $t_S^{\mu\nu;\rho\sigma}(q)$. Hence the latter corresponds
to the part coming from the traces of the energy-momentum tensors.
The Wilson coefficients in eq.~\re{TTexp1} are then of the general form
\be \begin{split}
C_{i}^{\mu\nu;\rho\sigma}(q)=&
\sum \limits_{r=1,5} \,t_r^{\mu\nu;\rho\sigma}(q)\,(Q^2)^{\f{-\text{dim}(O_i)}{2}}\,C_{i}^{(r)}(Q^2)\\
=&
\sum \limits_{r=T,S} \,t_r^{\mu\nu;\rho\sigma}(q)\,(Q^2)^{\f{-\text{dim}(O_i)}{2}}\,C_{i}^{(r)}(Q^2) \;\;(+\text{ contact terms)}.  
    \end{split} \label{Ci2ten} \ee
where $\text{dim}(O_0)=0$ and $\text{dim}(O_1)=4$ are the mass dimensions of the respective operators. \\
The coefficients defined in the first line of eq.~\re{Ci2ten} and their conversion to the ones defined in the second line
are given in the appendix.

In order to compute the coefficient $C_1^{\mu\nu;\rho\sigma}(q)$ we have used
the method of projectors \cite{Gorishnii:1983su,Gorishnii:1986gn} which allows to
express coefficient functions for any OPE of two operators in terms of
massless propagator type diagrams only. 
The method is based on dimensional regularization and uses strongly the fact that this
regularization sets every massless tadpole-like Feynman integral to zero.\\
The idea is to apply the same projector to both sides of eq.~\re{TTexp2_B} or to eq.~\re{TTexp2_R}
after contracting the free Lorentz indices with a tensor $\tilde{t}^{(r)}_{\mu\nu;\rho\sigma}(q)$ composed of the momentum  
$q$ and the metric $g$ in order to get the scalar pieces in eq.~\re{Ci2ten}:
\be 
\begin{split}
& {\bf P}\{\tilde{t}^{(r)}_{\mu\nu;\rho\sigma}(q)T^{\mu\nu;\rho\sigma}(q)\}=\sum \limits_i C_{i}^{B,(r)}(Q^2)\, {\bf P}\{O_{i}^B\},\\
& {\bf \hat{P}}\{\tilde{t}^{(r)}_{\mu\nu;\rho\sigma}(q)T^{\mu\nu;\rho\sigma}(q)\}=\sum \limits_i C_{i}^{(r)}(Q^2)\, {\bf \hat{P}}\{[O_{i}]\},\\
\end{split}
 \label{TTexp3}
\ee
The projector ${\bf P}$ or ${\bf\hat{P}}$
is constructed in such a way that it maps every operator on the rhs of the OPE to zero except for the one whose Wilson
coefficient we want to compute. The $\tilde{t}^{(r)}_{\mu\nu;\rho\sigma}$ can be constructed as linear combinations of the
$t_r^{\mu\nu;\rho\sigma}(q)$ from the list \re{t1_t5}.

As an example of how the method of projectors works let us consider the scalar object
\beq
\hat{T}_4(q) \equiv \frac{t_4^{\mu\nu;\rho\sigma}}{Q^4}  \hat{T}^{\mu\nu;\rho\sigma}(q)
\label{T4}
\eeq
which  should meet the following OPE
\be \hat{T}_4(q) 
\bbuildrel{=\!=\!=}_{q^2 \to -\infty}^{} 
c_{4,0}(q) \mathds{1} + c_{4,1}(q) [O_1]  +    {\cal{O}}(1/(-q^2)),
\label{ope:c4}\ee
with 
\beq	
c_{4,0}(q) = \sum_{r=1,5} g_{\mu\nu}\,g_{\rho\sigma}\,  t_r^{\mu\nu;\rho\sigma} C_0^{(r)}/Q^4
= C_0^{(1)} + 2 D C_0^{(2)} + 4  D C_0^{(3)} +  D^2 C_0^{(4)} +2 D C_0^{(5)}
\eeq
plus a similar equation relating $c_{4,1}(q)$ and $C_1^{(r)}$. 
Clearly, this procedure can be repeated in order to find the combinations $c_{1,0},c_{2,0},c_{3,0}$ and $c_{5,0}$ 
as well as  $c_{1,1},c_{2,1},c_{3,1}$ and $c_{5,1}$ corresponding to 
the use of the remaining four kinematical structures from the list \re{t1_t5} instead of
$t_4^{\mu\nu;\rho\sigma}$ in \re{T4}.
In order to extract the coeffcient function $c_{4,1}$ it is natural 
to consider the following (connected) Green function
\beq
T_4(k_1,k_2,q)  = 
\int\!\!\!\!\int\!\!\!\!\int\!\mathrm{d}^4x\,\mathrm{d}^4y_1\,\mathrm{d}^4y_2\,e^{iqx + k_1 y_1 + k_2 y_2}\, 
\langle 0 |
T[A^a_{\mu}(y_1)\, A^a_{\mu}(y_2)\,
 \hat{T}_4(x)
]
| 0\rangle^{\mathrm{amp}}
{},
\label{T4_2}
\eeq	
with the upperscript $\langle \dots \rangle^\mathrm{amp}$ meaning that we do not consider self-energy corrections to the
external gluon legs. Figure \re{O1_dias} shows some sample diagrams contributing to eq.~\re{T4_2}
at tree and one-loop level. 

As a  consequence of   eq.~\re{ope:c4} the Green function $T_4$ meets an OPE:
\beq
T_4(k_1,k_2,q) \bbuildrel{=\!=\!=}_{q^2 \to -\infty}^{}  c_{4,1}(q) 
\langle k_1 |[O]_1 | k_2\rangle>  + {\cal{O}}(1/(-q^2))
{},
\label{k1k2T4}
\eeq
with 
\beq
\langle k_1 |[O]_1 | k_2\rangle = 
\int\!\!\!\!\int\!\mathrm{d}^4y_1\,\mathrm{d}^4y_2\,e^{ k_1 y_1 + k_2 y_2}\, 
\langle 0 |
T[A^a_{\mu}(y_1)\, A^a_{\mu}(y_2)
 [O_1](0)
]
| 0\rangle^{\mathrm{amp}}
{}.
\label{k1k2O1}
\eeq

The ``standard'' way of computing $c_{4,1}$ from eq.~\re{k1k2T4} would be as follows:
\begin{itemize}

\item compute the  large Q asymptotic of $T_4(k_1,k_2,q)$ up to  and including  all terms which are not power-suppressed;

\item compute the matrix element $\langle k_1 |[O]_1 | k_2\rangle$;

\item  find $c_{4,1}$ by dividing out  the matrix element $ \langle  k_1 |[O]_1 | k_2\rangle$ from
large Q asymptotic of $T_4(k_1,k_2,q)$.
\end{itemize}
Note that in this  approach one could directly work with renormalized quantities  and with the  space-time 
dimension $D = 4 -2\epsilon$ set to its physical value 4.

The idea of the method of projectors is\footnote{
We discuss  below a kind of informal introduction to the approach; the reader  could find a more formal
exposition in \cite{Smirnov:2002pj}.}
to keep Q fixed but, instead,
consider the two momenta $k_1$ and $k_2$ of spectator gluons as {\em
infinitesimally} small and expand naively\footnote{ By naive expansion
we mean that for every contributing Feynman integral one expands the
corresponding Feynman {\em integrand} in the Taylor series in $k_i$
before any loop integrations are performed.
This also assumes that we do not put $D=4$  until the coefficient function(s) we are looking  for
are found; see below.}  both functions
$T_4(k_1,k_2,q)$ as well as $\langle k_1 |[O]_1 | k_2\rangle$ in these
momenta up to the second order.  As a result of this prescription:
\begin{itemize}

\item the  Green function $T_4(k_1,k_2,q)$ will loose its dynamical dependence on the momenta  $k_i$ and will constitute
of propagator-like diagrams;

\item   the matrix element $\langle k_1 |[O]_1 | k_2\rangle$ will only get contributions from tree diagrams
as all non-tree diagrams, being massless tadpoles,  are set to zero in dimensional regularization;

\item the  power-suppressed terms (in $q$) will dissappear (they contribute only to higher then  quadratic terms
of the Taylor expansion in $k_i$).

\item  both $T_4(k_1,k_2,q)$ and $\langle k_1 |[O]_1 | k_2\rangle$ will stop  to be finite
(we assume that we start from UV renormalized quantities) due to severe  IR divergences induced by
the naive expansion procedure. These divergences will be dimensionally regularized and manifest themselves 
as poles in $\epsilon$. The most important fact is that the coeffcient function $c_{4,1}$, being independent of
the small momenta $k_i$, will survive the procedure untouched! 
\end{itemize}

The tree level matrix element of the (renormalized) operator $[O_1]$ is given by the expression
\beq
\langle k_1 |[O]_1 | k_2\rangle = Z_{G}\, Z_3 \, n_g\,(D-1)\, k_1\cdot k_2
{},
\label{O1_tree}
\eeq
where $Z_3$ is the gluon wave  function  renormalization constant.
As a result we arrive at:
\beq
c_{4,1} =  \frac{1}{ Z_{G}\, Z_3}\, \hat{P}( T_4(k_1,k_2,q))
\label{P1}
{},
\eeq
where 
\be \hat{P}\Bigl( \dots \Bigr)=\f{\delta^{ab}}{\Ng}\f{g^{\mu_1 \mu_2}}{(D-1)}
\f{1}{D}\f{\p}{\p k_1} \cd \f{\p}{\p k_2} \Bigl( \dots \Bigr)|_{{}_{\scriptstyle k_1=0,k_2=0}}
\label{p1}
{}.
\ee
The explicit formula \re{P1} directly expresses the coefficient function to be found in terms of 
one-scale propagator-like  integrals.

Following the same logic we can construct a projector on the bare coefficient $C_{1,B}^{\mu\nu;\rho\sigma}(q)$.
Graphically it can be written as
\be
C_{1,B}^{\mu\nu;\rho\sigma}(q) =\f{\delta^{ab}}{\Ng}\f{g^{\mu_1 \mu_2}}{(D-1)}
\f{1}{D}\f{\p}{\p k_1} \cd \f{\p}{\p k_2} \left. \left[
  \begin{picture}(165,50) (0,0)
    \SetWidth{0.5}
    \SetColor{Black}
    \Gluon(10,0)(50,0){5.5}{4.5}
    \Gluon(110,0)(150,0){5.5}{4.5}
    \LongArrow(35,15)(25,15)
    \LongArrow(125,15)(135,15)
\DashCArc(80,0)(56,45,135){4}
\Photon(80,0)(120,40){3}{4}
\Photon(80,0)(40,40){3}{4}
    \CCirc(80,0){30}{Black}{Blue}
    \SetColor{Red}
\Vertex(110,0){4}
\Vertex(50,0){4}
    \SetColor{Black}
    \Text(25,17)[lb]{\Large{\Black{$k_1$}}}
    \Text(125,17)[lb]{\Large{\Black{$k_2$}}}
    \Text(36,-17)[lb]{\Large{\Red{$g_B$}}}
    \Text(110,-17)[lb]{\Large{\Red{$g_B$}}}
    \Text(150,-20)[lb]{\Large{\Black{$\mu_2$}}}
    \Text(5,-20)[lb]{\Large{\Black{$\mu_1$}}}
    \Text(5,10)[lb]{\Large{\Black{$a$}}}
    \Text(150,10)[lb]{\Large{\Black{$b$}}}
    \Text(27,40)[lb]{\Large{\Black{$\mu\nu$}}}
    \Text(122,40)[lb]{\Large{\Black{$\rho\sigma$}}}
  \end{picture}
\right] \right|_{k_i=0}
{},
\ee
where the blue circle represents the the sum of all (bare) Feynman diagrams
which become 1PI after formal gluing (depicted as a dotted line above) of
the two external lines representing the operators on the lhs of the OPE and
carrying the (large) momentum q. The use of $g_B=Z_g g_s$ on the vertices  at the end
of the external gluon lines already contains the renormalization factor $\f{1}{Z_3}$ from eq.~\re{P1}.
The renormalized coefficient is then derived according to eq.~\re{C1ren}. 
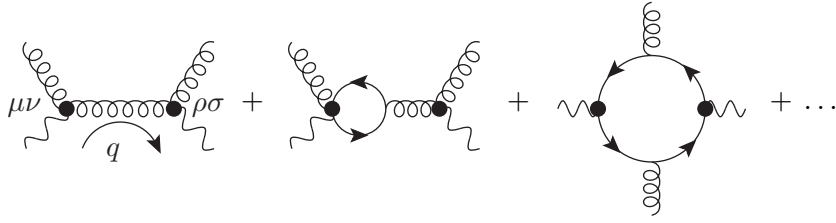
\begin{figure}[h!] 
\begin{center}
\vskip 2ex
$  \begin{picture}(80,30) (0,0)
    \SetWidth{0.5}
    \SetColor{Black}
\Photon(5,-15)(20,0){3}{2}
\Photon(60,0)(75,-15){3}{2}
\Gluon(20,0)(60,0){3}{6.5}
\Gluon(20,0)(5,25){3}{4.5}
\Gluon(60,0)(75,25){3}{4.5}
\LongArrowArcn(40,-20)(15,160,25)
    \Vertex(20,0){3}
    \Vertex(60,0){3}
    \Text(-2,-3)[lb]{{\Black{$\mu \nu$}}}
    \Text(67,-3)[lb]{{\Black{$\rho \sigma$}}}
    \Text(35,-20)[lb]{{\Black{$q$}}}
  \end{picture}   \; + \;
\begin{picture}(80,30) (0,0)
    \SetWidth{0.5}
    \SetColor{Black}
\Photon(5,-15)(20,0){3}{2}
\Photon(60,0)(75,-15){3}{2}
\Gluon(40,0)(60,0){3}{3.5}
\ArrowArc(30,0)(10,180,0)
\ArrowArc(30,0)(10,0,180)
\Gluon(20,0)(5,25){3}{4.5}
\Gluon(60,0)(75,25){3}{4.5}
    \Vertex(20,0){3}
    \Vertex(60,0){3}
  \end{picture}\; + \;
\begin{picture}(80,30) (0,0)
    \SetWidth{0.5}
    \SetColor{Black}
\Photon(5,0)(20,0){3}{2}
\Photon(60,0)(75,0){3}{2}
\ArrowArc(40,0)(20,180,270)
\ArrowArc(40,0)(20,90,180)
\ArrowArc(40,0)(20,0,90)
\ArrowArc(40,0)(20,270,360)
\Gluon(40,20)(40,40){3}{3.5}
\Gluon(40,-20)(40,-40){3}{3.5}
    \Vertex(20,0){3}
    \Vertex(60,0){3}
  \end{picture}\,+\,\ldots$\\[9ex]
\caption{Diagrams for the calculation of the coefficient $C_1$. \label{O1_dias}}
\end{center}
\end{figure}

\ice{

It is interesting to see that at tree level only contact terms are produced and therefore
the lowest physically important order is $\als$ (one loop).
For matrix elements with physical external states, e.g. the vacuum, the renormalized operator $[O_1]$ is derived from the bare operator
$O_1^B$, where all gluon fields are the bare ones, by a simple multiplicative renormalization.\footnote{This is only true in 
massless QCD and for physical external states. In general $O_1^B$ mixes with unphysical operators and 
operators proportional to the quark masses under renormalization (see \cite{Spiridonov:1984br}).}
}

\section{Results}
\ice{
In physical applications like sum rules we are only interested in the scale dependent part of the Wilson coefficients, i.e.
parts which depend on powers of $\f{1}{Q^2}$ or $l_{\sss{\mu q}}:=\ln\lb\f{\mu^2}{Q^2}\rb$. For simplicity and unambiguousness
we therefore present the Adler functions of the scalar pieces in $C_0^{\mu\nu\rho\sigma}$. 
}

All results are given in the $\overline{\text{MS}}$ scheme with
$\as=\f{\als}{\pi}$, $\als=\f{\gs^2}{4\pi}$ and the
abbreviation $l_{\sss{\mu q}}=\ln\lb\f{\mu^2}{Q^2}\rb$ where $\mu$ is the
$\overline{\text{MS}}$ renormalization scale.
They can be retrieved from\\
\texttt{\bf http://www-ttp.particle.uni-karlsruhe.de/Progdata/ttp12/ttp12-025/}

The gauge group factors are defined in the usual way: $\cf$ and $\ca$ are the quadratic Casimir
operators of the quark and the adjoint representation of the corresponding Lie algebra,
$\dR$ is the dimension of the quark representation, $\Ng$ is the number of gluons (dimension of the adjoint representation),
$\tr$ is defined so that \mbox{$\tr \delta^{ab}=\textbf{Tr}\lb T^a T^b\rb$}  
is the trace of two group generators of the quark representation.\footnote{For an SU$(N)$ gauge group these are $\dR=N$,
$\ca=2\tr N$ and $\cf=\tr\lb N-\f{1}{N}\rb$.}
For  QCD (colour gauge group SU$(3)$) we have $\cf =4/3\,,\, \ca=3\,,\,\tr=1/2$ and $\dR = 3$.
By $\Nf$ we denote the number of active quark flavours.

\subsection{$C_0$}

Because of the contact terms both coefficients $C_0^{S} $ and $C_0^{T}$ could be unambiguously computed only up to constant
(that is q-independent) contributions. To avoid  the ambiguity we present below their  $Q^2$-derivatives:
\be \begin{split}
Q^2\f{d}{dQ^2}\,C_0^{(T)}=&
       \f{1}{16\pi^2}\left[
          - \f{1}{10} \Ng
          - \f{1}{20} \Nf \dR \right.\\
       &+ \as   \left\{
           \f{1}{18} \ca \Ng
          - \f{7}{144} \Nf \tr \Ng
          \right\}\\
       &+ \as^2   \left\{
           \f{67}{12960} \ca^2 \Ng
          + \f{3}{128} \Nf \tr \cf \Ng
          - \f{10663}{51840} \Nf \ca \tr \Ng
          + \f{473}{6480} \Nf^2 \tr^2 \Ng  \right. \\&\;\;\,\qquad
          + \f{11}{216} l_{\sss{\mu q}} \ca^2 \Ng
          - \f{109}{1728} l_{\sss{\mu q}} \Nf \ca \tr \Ng
          + \f{7}{432} l_{\sss{\mu q}} \Nf^2 \tr^2 \Ng  \\&\;\;\,\qquad\left.\left.
          + \f{11}{40} \zeta_{3} \ca^2 \Ng
          + \f{3}{80} \zeta_{3} \Nf \ca \tr \Ng
          - \f{1}{20} \zeta_{3} \Nf^2 \tr^2 \Ng
          \right\}\right].
\end{split} \label{adlerC0T}\ee
\vskip 2ex
\be \begin{split}
Q^2\f{d}{dQ^2}\,C_0^{(S)}(Q^2)=&
       \f{\as^2}{16\pi^2}   \left\{
          - \f{121}{1296} \ca^2 \Ng
          + \f{11}{162} \Nf \ca \tr \Ng
          - \f{1}{81} \Nf^2 \tr^2 \Ng
          \right\}
\\
 =& -\frac{a_s^2}{144\,\pi^2}  \beta_0^2\,n_g
{},
\end{split} \label{adlerC0S}
\ee
where 
\[\beta_0 = \frac{11\, C_A}{12}  - \frac{n_f\,T_f}{3}
\]
is  the first coefficient  of the perturbative  expansion of the  $\beta$-function
\re{be:def}.
\ice{
\beq
Q^2\frac{d}{d Q^2}\,C_0^{(S)}(Q^2) = -\frac{a_s^2}{16\,\pi^2}  \beta_0^2\,n_g
\eeq
}
This result for $Q^2\f{d}{dQ^2}\,C_0^{(T)}$ is in agreement with the 
one derived in \cite{Pivovarov_tensorcurrents}
for the case of gluodynamics ($n_f=0$) at order $\als$ (two-loop level).
The simple form of eq.~\re{adlerC0S} comes from  
the  well-known trace anomaly  \cite{Collins:1976yq,nielsen_Tmunu}, which reads
\be \label{Tmumu_G2}
T^\mu_{\,\,\mu}=\f{\beta(a_s)}{2} \, [G^a_{\rho\sigma} G^{a\,\rho\sigma}]
= -2\,\beta(a_s) \, [O_1]
{}.
\ee
Indeed, from {\em operator}  eq.~\re{Tmumu_G2} we expect that
\beq
i\int \! \mathrm{d}^4x\,e^{iqx}\,
\langle 0|T[\,T^\mu_{\,\,\mu}(x)T^\nu_{\,\,\nu}(0)]|0\rangle
= 4\, \beta^2(\als) \,Q^4\, \Pi^{GG}(q^2) + \mbox{contact terms},
\label{expect}
\eeq
where
\beq
Q^4\, \Pi^{GG}(q^2) = i\int \! \mathrm{d}^4x\,e^{iqx}\,
\langle 0|T\,[O_1 (x) O_1(0)]|0\rangle
\label{O1O1:def}
{}.
\eeq
Now the  one-loop  result
\be
Q^2\f{d}{dQ^2}\,\Pi^{GG}(q^2) =\quad
  \begin{picture}(100,40) (0,0)
    \SetWidth{0.5}
    \SetColor{Black}
\Photon(0,0)(20,0){3}{2.5}
\Photon(80,0)(100,0){3}{2.5}
\LongArrow(2,8)(12,8)
\LongArrow(85,8)(95,8)
\GlueArc(50,0)(30,0,180){3}{9.5}
\GlueArc(50,0)(30,180,360){3}{9.5}
    \SetColor{Black}
    \Vertex(20,0){3}
    \Vertex(80,0){3}
    \Text(5,-14)[lb]{{\scriptsize \Black{$O_1$}}}
    \Text(86,-14)[lb]{{\scriptsize \Black{$O_1$}}}
    \Text(3,12)[lb]{{\Black{$q$}}}
    \Text(85,12)[lb]{{\Black{$q$}}}
  \end{picture}\quad =-\f{1}{64\pi^2}\, \Ng +  \mbox{contact terms}\\[8ex]
\ee
leads directly to eq.~\re{adlerC0S}
\ice{
and the $\beta$-function
\be \begin{split} \beta(\als)=&\mu^2 \f{d}{d\mu^2}\als(\mu^2) 
           = \f{\als^2}{4\pi} \left\{ - \f{11}{3} \ca + \f{4}{3} \Nf\tr\right\} 
           . \label{betaals} \end{split} \ee
}
The fact that this particular three-loop result can be derived from one-loop results is also
the reason for the lack of $\zeta$-functions in it. Furthermore the structure of eq.~\re{Tmumu_G2} explains
nicely why the leading contribution for this scalar piece is of order $\als^2$.

In fact, the correlator \re{O1O1:def} is known in two-, three- and four-loop approximations from works 
\cite{Kataev:1982gr},\cite{Chetyrkin:1997iv} and \cite{Baikov:2006ch} respectively. The four-loop result
reads (with all colour factors set to their QCD values and $l_{\mu q} =0$)
\bea
Q^2\f{d}{dQ^2}\,\Pi^{GG}(q^2) = \f{1}{16\pi^2}\Biggl\{
&{-}&2
+a_s\,\Biggl(
-\frac{73}{2}
+ 
 \frac{7}{3}\,n_f 
\Biggr)
\nonumber
\\
&+&
a_s^2\,\Biggl(
-\frac{37631}{48} 
+\frac{495}{4}  \zeta_{3}
+
n_f 
\left[
\frac{7189}{72} 
-\frac{5}{2}   \zeta_{3}
\right]
- 
  \frac{127}{54}  n_f^2
          \Biggr)
\nonumber
\\
&{+}&
a_s^3\,
\Biggl(
-\frac{15420961}{864} 
+\frac{44539}{8}  \sbz \zeta_{3}
-\frac{3465}{4}  \sbz \zeta_{5}
\nonumber\\
&{+}& \,n_f 
\left[
\frac{368203}{108} 
-\frac{11677}{24}  \sbz \zeta_{3}
+\frac{95}{18}  \sbz \zeta_{5}
\right]
\nonumber\\
&{+}& \, n_f^2
\left[
-\frac{115207}{648} 
+\frac{113}{12}  \sbz \zeta_{3}
\right]
{+} \, n_f^3
\left[
\frac{7127}{2916} 
-\frac{2}{27}  \sbz \zeta_{3}
\right]
\Biggr)
\Biggr\}
\label{O1O:4loop}
{}.
\eea
Finally, using eq.~\re{expect} and the well-known result for the four-loop QCD $\beta$-function 
\cite{vanRitbergen:1997va,Czakon:2004bu} we could easily extend the rhs of \re{adlerC0S} by
{\em three more orders} in $\alpha_s$:
\bea
&{}& Q^2\f{d}{dQ^2}\,C_0^{(S)}(Q^2)=
 \frac{a_s^2}{16\,\pi^2} \Biggr\{
-\frac{121}{18} + \frac{22}{27} n_f
-\frac{2}{81}\, n_f^2
\nonumber
\\
&+& a_s \,
\Biggl(
-\frac{11077}{72} +  \frac{1025}{36}\, n_f
-\frac{265}{162}\, n_f^2 +  \frac{7}{243}\, n_f^2
\Biggr)
\nonumber
\\
&+&
a_s^2 \,
\Biggl(
-\frac{5787209}{1728} 
+\frac{6655}{16}  \sbz \zeta_{3}
{+} \,n_f 
\left[
\frac{540049}{648} 
-\frac{4235}{72}  \sbz \zeta_{3}
\right]
+
 \, n_f^2
\left[
-\frac{556555}{7776} 
+\frac{275}{108}  \sbz \zeta_{3}
\right] 
\nonumber
\\
&+& 
\hspace{2cm}
\, n_f^3
\left[
\frac{29071}{11664} 
-\frac{5}{162}  \sbz \zeta_{3}
\right] 
-\frac{127}{4374}
\Biggr)
\nonumber
\\
&+&
a_s^3 \,
\Biggl(
-\frac{2351076745}{31104} 
+\frac{5925007}{288}  \sbz \zeta_{3}
-\frac{46585}{16}  \sbz \zeta_{5}
\nonumber
\\
&+& \,n_f 
\left[
\frac{367411229}{15552} 
-\frac{33359777}{7776}  \sbz \zeta_{3}
+\frac{240185}{648}  \sbz \zeta_{5}
\right]
\nonumber\\
&{+}&\hspace{2cm} \, n_f^2
\left[
-\frac{381988321}{139968} 
+\frac{3715127}{11664}  \sbz \zeta_{3}
-\frac{12485}{972}  \sbz \zeta_{5}
\right]
\nonumber
\\
&+& 
\, n_f^3
\left[
\frac{20279497}{139968} 
-\frac{180083}{17496}  \sbz \zeta_{3}
+\frac{95}{1458}  \sbz \zeta_{5}
\right]
\nonumber\\
&{}&\hspace{2cm}{+}
 n_f^4
\left[
-\frac{1101389}{314928} 
+\frac{427}{2916}  \sbz \zeta_{3}
\right]
{+} \, n_f^5
\left[
\frac{7127}{236196} 
-\frac{2}{2187}  \sbz \zeta_{3}
\right]
\Biggr)
\Biggr\}
{}.
\eea

\subsection{$C_1$}
According to the definition of $C_1^{\mu\nu;\rho\sigma}$ in eq.~\re{Ci2ten} there is a factor $\f{1}{(Q^2)^2}$ in front of
the dimensionless scalar pieces $C_{1}^{(S)}$ and $C_{1}^{(T)}$ which makes the whole coefficient 
immune to contact terms except for those proportional to the tensor structures
$t_4^{\mu\nu;\rho\sigma}(q)$ and $t_5^{\mu\nu;\rho\sigma}$ defined in eq.~(\ref{t1_t5}). The
physical pieces $C_{1}^{(S)}$ and $C_{1}^{(T)}$ however are unambigous and the results read:
\bea
   C_{1}^{(S)} =&{}& \as\left\{
        \;\;\,\f{22}{27}   \ca
          - \f{8}{27}   \Nf \tr\right\}
\nonumber
\\ 
        &+& \as^2 \left\{
          \f{83}{324}\ca^2
          - \f{2}{9}\Nf\tr\cf
          - \f{8}{81}\Nf\ca\tr
          - \f{4}{81}\Nf^2\tr^2          \right\},
 \label{C1Sfin}
\\
   C_{1}^{(T)} = &{}&\as \left\{
       - \f{5}{18}   \ca
          - \f{5}{72}   \Nf \tr\right\}
\nonumber
\\
       &+& \as^2 \left\{
          - \f{83}{432}\ca^2
          + \f{43}{96}\Nf\tr\cf
          + \f{41}{432}\Nf\ca\tr
          - \f{1}{216}\Nf^2\tr^2          \right\}
 \label{C1Tfin}
{}.
\eea
\vskip 1ex
One thing to notice about  $C_{1}^{(S)}$ is that if we take
the trace of both energy-momentum tensors the whole term 
$\eta^\mu_{\,\,\mu}(q)\eta^\nu_{\,\,\nu}(q)\f{1}{(Q^2)^2}C_{1}^{(S)}$
in the Wilson coefficient becomes local and, therefore,  indistinguishable from contact terms.
\ice{
In fact this is necessary because from eq.~\re{Tmumu_G2} we expect
that the correlator of the traces of two energy-momentum tensors has no non-local contributions to the coefficient in front of $O_1$
until order $\als^2$.\footnote{Therefore the result for $C_{1}^{(S)}$ is only physically meaningful for $\mu \neq \nu$ or
$\rho \neq \sigma$.}
} 
We can however check eq.~\re{C1Sfin} independently by computing first the coefficient function  $C^{(TG,T)}_{1}$ in an OPE
( $t^{\mu\nu}_S = q^4\, g_{\mu\nu}, \ t^{\mu\nu}_T = q^4\, g_{\mu\nu} - q^2 q_{\mu}\, q_{\nu} $)
\be \begin{split}
& i\int \! \mathrm{d}^4x\,e^{iqx}\,T[\,T^{\mu\nu}\! (x) 
 G_{\rho\sigma}^2\!(0)]
 \bbuildrel{=\!=\!=}_{q^2 \to -\infty}^{}\\
& \hspace{2cm}\left( C^{(TG,S)}_{0}  t^{\mu\nu}_S +  C^{(TG,T)}_{0} t^{\mu\nu}_T\right) \mathds{1} 
+   \left( C^{(TG,S)}_{1} t^{\mu\nu}_S  + C^{(TG,T)}_{1} t^{\mu\nu}_T\right) \frac{[O_1]}{Q^4} +\ldots \end{split}
\ee
and then employing eq.~\re{Tmumu_G2} to get the next higher order in $\als$ for $C_{1}^{(S)}$.
The result 
\be C^{(TG,T)}_{1} = 
-\f{16}{3}+ \as\left(\f{22}{9}\ca - \f{8}{9}\Nf\tr\right) + {\cal O}(\als^2) 
= -\frac{16}{3}\, \left(1+ \beta(a_s)/2\right) + {\cal O}(\als^2)
\ee
allows to represent the rhs of eq.~\re{C1Sfin} in a   form directly confirming eq.~\re{Tmumu_G2}:
\beq
C_1^{(S)} = \frac{\beta(\as)}{6}\, C^{(TG,T)}_{1} +  {\cal O}(\als^3)
= -\frac{8}{9}\,\beta(\as)\, \left(1+ \beta(a_s)/2\right) + {\cal O}(\als^3)
{}.
\eeq
The factor $\beta(\as)$ in this result is a direct consequence of the trace anomaly equation \re{Tmumu_G2}.
However, we do not know any rationale behind the peculiar structure  after this factor.
If it is not accidental, then one  can hope that an explanation  could  be found within the so-called
$\beta$-expansion formalism suggested in \cite{Mikhailov:2004iq}.

It is important to note that the coefficient functions $C_{1}^{(S)}$ and $C_{1}^{(T)}$
are \textit{not} Renormalization Group independent. We can construct the corresponding RG invariants  by
using the well-known fact\footnote{This follows directly from the RG invariance
of the energy-momentum tensor and the trace anomaly equation
\re{Tmumu_G2}.} that the scale invariant version of the operator $O_1$
is 
\beq
O_1^{RGI} \equiv \hat{\beta}(a_s) \, [O_1], \ \ \  \hat{\beta}(a_s) = \frac{-\beta(a_s)}{\beta_0} = 
a_s\left(1+  \sum_{i \ge 1} \frac{\beta_i}{\beta_0} a_s^i \right)
{}. \label{O1RGI}
\eeq
From this and the scale invariance of 
$T^{\mu\nu;\rho\sigma}(q)$ defined in eq.~\re{TT} we find the RG invariant Wilson coefficients
\beq \begin{split}
C^{(S)}_{1,RGI} &\equiv C^{(S)}_{1}/\hat{\beta}(a_s)\\
C^{(T)}_{1,RGI} &\equiv C^{(T)}_{1}/\hat{\beta}(a_s)
\label{C1STRGI}
\end{split} \eeq
which satisfy 
\be C^{(S,T)}_{1,RGI} O_1^{RGI}=C^{(S,T)}_{1}[O_1]. \ee
From this definition we can immediately explain the absence of $l_{\mu q}$ in eq.~\re{C1Sfin} and eq.~\re{C1Tfin}.
Suppose we had $l_{\mu q}$ in $C^{(S)}_{1}$ and therefore in $C^{(S)}_{1,RGI}$ then the general structure
of eq.~\re{C1STRGI} up to three-loop order would be
\be \begin{split}
C^{(S,T)}_{1,RGI}&=(a_1+b_1\,l_{\mu q})+a_s(a_2+b_2\,l_{\mu q}+c_2\,l_{\mu q}^2)\\
&+a_s^2(a_3+b_3\,l_{\mu q}+c_3\,l_{\mu q}^2+d_3\,l_{\mu q}^3)
+\mathcal{O}(a_s^3) \end{split}
\ee
with scale independent coefficients $a_i,b_i,c_i$ and $d_i$.
The derivative with respect to $\mu^2$ must vanish:
\be \begin{split}
\mu^2\f{d}{d\mu^2} C^{(S,T)}_{1,RGI}&=b_1+a_s (b_2+ 2c_2\, l_{\mu q})+a_s\beta(a_s)(a_2+b_2\,l_{\mu q}+c_2\,l_{\mu q}^2)\\
&+ a_s^2(b_3+2c_3\,l_{\mu q} + 3d_3\,l_{\mu q}^2)
+\mathcal{O}(a_s^3)\bbuildrel{=}_{}^{!} 0 \quad \forall \,\mu^2\\
&\Rightarrow b_1=0\\
&\Rightarrow b_2=0,\, c_2=0\\
&\Rightarrow b_3=\beta_0 a_2,\,c_3=0,\,d_3=0.\\
    \end{split}
\label{RG:constraints}
\ee
In conclusion, not only have we explained the absence of logarithms in eq.~\re{C1Sfin} and eq.~\re{C1Tfin} 
but we also get the logarithmic part of the three-loop result for these coefficient functions for free.
Terms with $l_{\mu q}^2$ can only appear starting from four-loop level, terms with $l_{\mu q}^3$
from five-loop level and so on.\\
The quantities defined in eq.~\re{C1STRGI} are given by
\bea 
C^{(S)}_{1,RGI}=&\f{22}{27} \ca-\f{8}{27} \Nf \tr-
\f{a_s}{324} \left(11\ca-4\Nf\tr\right)^2\\
C^{(T)}_{1,RGI}=&-\f{5}{72} (4 \ca+\Nf \tr)+\f{a_s}{864 (11 \ca-4 \Nf \tr)}
 \left(214 \ca^3+876 \ca^2 \Nf \tr \right. \nonumber \\ 
&\left.+3537 \ca \cf \Nf \tr-672 \ca \Nf^2 \tr^2-1728 \cf \Nf^2 \tr^2+16 \Nf^3 \tr^3\right)
\eea
The three-loop parts proportional to $l_{\mu q}$ are
\bea 
C^{(S,\text{3l,log})}_{1,RGI}=&a_s^2l_{\mu q}\left\{-\f{1331 \ca^3}{3888}+\f{121}{324} \ca^2 \Nf \tr
-\f{11}{81} \ca \Nf^2 \tr^2+\f{4}{243} \Nf^3 \tr^3\right\}
\label{CS:3l:log}
{},
\\
C^{(T,\text{3l,log})}_{1,RGI}=&a_s^2l_{\mu q}\left\{
\f{214 \ca^3+876 \ca^2 \Nf \tr +3537 \ca \cf \Nf \tr -672 \ca \Nf^2 \tr^2-1728 \cf \Nf^2 \tr^2+16 \Nf^3 \tr^3}{10368}\right\}
\label{CT:3l:log}
{}.
\eea
For completeness we  have also computed the contribution of the gluon condensate to the OPE of 
correlator  \re{O1O1:def}:
\beq
 Q^4\, \Pi^{GG}(q^2) \bbuildrel{=\!=\!=}_{q^2 \to -\infty}^{}   C_0^{GG} \,Q^4 
\ \ + \ \  C_1^{GG} \, \langle 0 |[O_1]| 0\rangle
\eeq
with the result:
\be 
\begin{split}
 C_1^{GG}= &-1
       + a_s   \left(
          - \f{49}{36} \ca
          + \f{5}{9} \Nf \tr
          - \f{11}{12} l_{\sss{\mu q}} \ca
          + \f{1}{3} l_{\sss{\mu q}} \Nf \tr
          \right)\\
       &+ a_s^2   \left(
          - \f{11509}{1296} \ca^2
          + \f{13}{4} \Nf \tr \cf
          + \f{3095}{648} \Nf \ca \tr
          - \f{25}{81} \Nf^2 \tr^2 \right. \\ &\left.
          - \f{1151}{216} l_{\sss{\mu q}} \ca^2 
          + l_{\sss{\mu q}} \Nf \tr \cf
          + \f{97}{27} l_{\sss{\mu q}} \Nf \ca \tr
          - \f{10}{27} l_{\sss{\mu q}} \Nf^2 \tr^2 \right. \\ &\left.
          - \f{121}{144} l_{\sss{\mu q}}^2 \ca^2
          + \f{11}{18} l_{\sss{\mu q}}^2 \Nf \ca \tr
          - \f{1}{9} l_{\sss{\mu q}}^2 \Nf^2 \tr^2
          + \f{33}{8} \zeta_{3} \ca^2  \right. \\ &\left.
          - 3 \zeta_{3} \Nf \tr \cf
          + \f{3}{2} \zeta_{3} \Nf \ca \tr
          \right)\\
      &+ \fbox{$
         \f{a_s^2}{\eps}  \left(
          - \f{17}{24} \ca^2
          + \f{1}{4} \Nf \tr \cf
          + \f{5}{12} \Nf \ca \tr
          \right)
          $}
          {}\,  .
\label{C1:GG}
\end{split}
\ee
The tree and one-loop contributions in \re{C1:GG} are in agreement with
\cite{Novikov_scalargluonium} and \cite{Bagan:1989vm,Harnett:2004pg}
correspondingly.  The two-loop part is new and has a feature that did
not occur in lower orders, namely, a divergent contact term. Its
appearance clearly demonstrates that non-logarithmic perturbative 
contributions to $C_1^{GG}$ are {\em not} well defined in  QCD, a fact seemingly ignored by
the QCD sum rules   practitioners (see, e.g. \cite{forkel_sumrule,Harnett:2000fy}).
It is an interesting to notice that this divergent term is equal to $-\f{a_s^2}{\eps} \beta_1$
(We thank M. Jamin for drawing our attention to this). This could point to the possibility
that the contact terms and therefore the missing part of a complete renormalization of $C_1$ could
be expressed in some way through the $\beta$-function. This remains an open problem for the moment.

An unambiguous QCD prediction can be made for the derivative:
\beq
\begin{split}
 Q^2\f{d}{dQ^2}\, C_1^{GG}= &
  a_s   \left(
           \f{11}{12}  \ca
          - \f{1}{3}  \Nf \tr
          \right)\\
       &+ a_s^2   \left(
           \f{1151}{216}  \ca^2 
          -  \Nf \tr \cf
          - \f{97}{27}  \Nf \ca \tr
          + \f{10}{27}  \Nf^2 \tr^2 \right. 
         \\ &\left.
          + \f{121}{72} l_{\sss{\mu q}} \ca^2
          - \f{11}{9} l_{\sss{\mu q}} \Nf \ca \tr
          + \f{2}{9} l_{\sss{\mu q}} \Nf^2 \tr^2
             \right)
          {}\,  .
\label{dC1:GG}
\end{split}
\eeq

\ice{For the case $\mu=\nu$ and $\rho=\sigma$ the $C_{1}^{(S)}$ in eq.~\re{C1Sfin} is not valid. 
The correct one should be determined
from eq.~\re{Tmumu_G2} using the results given in
\cite{Novikov_scalargluonium, Novikov1979347}.
}

\section{Numerics}

In this section we will give our main results in the numerical form
for two cases of interest, that is gluodynamics ($n_f=0$) and QCD with
three light quarks only ($n_f=3$). As has already been mentioned, not
all coefficient functions which we have discussed in the previous section
are Renormalization Group independent. For a meaningful 
discussion we will construct the corresponding RG invariants  by
using the scale invariant version of the operator $O_1$ defined in eq.~\re{O1RGI}.
In addition we
set $l_{\sss{\mu q}} = 0$ everywhere.\footnote{This corresponds to the choice $\mu^2=Q^2$ for the renormalization scale.} 
\beq
Q^2\f{d}{dQ^2}\,C_0^{(T)} \bbuildrel{=\!=\!=}_{n_f=0}^{}  -\f{4}{80\pi^2}
\left(1 - 1.66667 \, a_s - 30.2162 \, a_s^2
\right
)
{},
\eeq
\beq
Q^2\f{d}{dQ^2}\,C_0^{(T)} \bbuildrel{=\!=\!=}_{n_f=3}^{}  -\f{5}{64\pi^2}
\left(1 -  0.6 \, a_s - 15.1983 \, a_s^2
\right
)
{},
\eeq
\beq
Q^2\f{d}{dQ^2}\,C_0^{(S)} \bbuildrel{=\!=\!=}_{n_f=0}^{}  
-\f{121}{288\pi^2}a_s^2\left(1+22.8864 a_s+423.833 a_s^2+8014.74 a_s^3\right)
{},
\eeq
\beq
Q^2\f{d}{dQ^2}\,C_0^{(S)} \bbuildrel{=\!=\!=}_{n_f=3}^{}  
-\f{9}{32\pi^2}a_s^2\left(1+18.3056 a_s+247.48 a_s^2+3386.41 a_s^3\right)
{},
\eeq
\beq
Q^2\f{d}{dQ^2}\,C_1^{GG,RGI}  \bbuildrel{=\!=\!=}_{n_f=0}^{}  \f{11}{4}a_s^2
\left(1   + 19.7576 \, a_s
\right ),  \ \ \ C_1^{GG,RGI}  \equiv \hat{\beta}( a_s) \, C_1^{GG}
{}, 
\eeq
\beq
Q^2\f{d}{dQ^2}\,C_1^{GG,RGI}  \bbuildrel{=\!=\!=}_{n_f=3}^{}  \f{9}{4}a_s^2
\left(1   + 15.3889 \, a_s
\right )
{}, 
\eeq
\beq
C^{(S)}_{1,RGI}  \bbuildrel{=\!=\!=}_{n_f=0}^{}  \frac{22}{9} \left( 1- 1.375 \, a_s \right),
\ \ \  C^{(S)}_{1,RGI} \equiv C^{(S)}_{1}/\hat{\beta}( a_s)
{},
\eeq
\beq
C^{(S)}_{1,RGI}  \bbuildrel{=\!=\!=}_{n_f=3}^{}  2 \left( 1 - 1.125  \, a_s\right)
{},
\eeq
\beq
C^{(T)}_{1,RGI}  \bbuildrel{=\!=\!=}_{n_f=0}^{}  -\frac{5}{6} \left( 1 -  0.2431825 \, a_s \right)
{},
\ \ \  C^{(T)}_{1,RGI} \equiv C^{(T)}_{1}/\hat{\beta}( a_s)
{},
\eeq
\beq
C^{(T)}_{1,RGI}  \bbuildrel{=\!=\!=}_{n_f=3}^{}  -\frac{15}{16} \left( 1 - 1.3333 \, a_s \right)
{}.
\eeq

\section{Applications to high-temperature QCD} Recently, the
correlators $\Pi^{GG}$ and $T^{\mu\nu;\rho\sigma}(q)$ have been studied
in \ice{ high-temperature} (Euclidean)  hot Yang-Mills theory in
\cite{Laine:2010tc,Schroder:2011ht} respectively (see, also references
therein for related earlier works).

In this section we will employ our $\bold T=0$ calculations 
in order to  extend {\em some} of the results of  these publications 
by adding fermionic  contributions as well as  higher 
order corrections. Note that for simplicity we will set all colour factors in all expressions below
to their QCD values. The reader interested in expressions  valid for generic  colour group should be able
to derive the corresponding results himself from our results.

\subsection{Trace anomaly correlator}

In ~\cite{Laine:2010tc}   two-loop corrections to   the quantity\footnote{Note that  $G_{\theta}( 0, \vec{X})$
has been  directly  measured  in lattice simulations \cite{Iqbal:2009xz}.} 
\beq
G_{\theta}(X) \equiv   \langle T[\ta(X)\, \ta(0) ] \rangle_c, \ \  \ \ta \equiv T^{\mu}_{\,\,\mu} 
\ice{=  -2\,\beta(a_s) \, O_1}
{},
\label{Gtheta}
\eeq
where $\langle \dots \rangle_c$ stands for  the connected part and the expectation value
is taken at finite temperature\footnote{We use the bold case for the temperature to make
it distinct from $T(\dots)$ standing for the time ordered product of operators inside the
round brackets.}       $\bold T$,   have been computed.
The  capital case $X$ for the space-time argument in \re{Gtheta} is used 
in order to stress that we  are dealing with a Euclidean correlator. In the following $e$ and $p=|\bold q|$
are the energy density and the pressure of the system with the well-known relation $\langle \theta \rangle_c=e-3p$. 
In the limit  of small $r \equiv |{X}|$  
the result of \cite{Laine:2010tc} \mbox{reads}
\footnote{The expression below is the somewhat modified eq.~(5.7) of \cite{Laine:2010tc}.}
\beq
 \frac{4\,a_s^2}{\beta^2(a_s)} G_{\theta}(r) 
  =  
 \frac{384}{\pi^4 r^8}
 \,\bar{\gamma}_\rmi{$\theta;\unit$}(r)
 \; -  \;
 \frac{8\,a_s\, \langle \theta \rangle_c}{\beta(a_s)\pi^2 r^4}
 \,\bar{\gamma}_\rmi{$\theta;\theta$}(r)
  \; - \;  
 \frac{64 (e+p)}{\pi^2 r^4}
 \,\bar{\gamma}_\rmi{$\theta;e+p$}(r)
\; + \; {\cal O}\biggl( \frac{{\bold{T}}^6}{r^2} \biggr) 
{},
\eeq
with
\bea
\bar{\gamma}_{\theta;\unit}(r)&=&a_s^2 + a_s^3\,\Biggl(
-\frac{1}{12} 
+\frac{11}{2}  \,l_{\mu X}\,
\Biggr) +{\mathcal O}(a_s^4)
{}, 
\label{gamma1}\\
\bar{\gamma}_{\theta;\theta}(r)&=&  22\, a_s^2 + {\mathcal O}(a_s^3) \label{gammae3p},\\
\bar{\gamma}_{\theta;e+p}(r)&=& a_s^2 +a_s^3\Biggl(
 \frac{15}{72}  + \frac{11}{2}\,l_{\mu X}\ 
\Biggr) 
+{\mathcal O}(a_s^4) 
{},
\eea
and $l_{\mu X}=\log(\mu^2X^2/4)+2\gamma_E$.

According to  \cite{CaronHuot:2009ns} the coefficient functions
$\bar{\gamma}_{\theta;\unit}$ and $\bar{\gamma}_{\theta;\theta}(r)$ do not depend
on temperature ${\bf T}$ and, thus, should coincide with their ${\bf T}=0$
counterparts. Hence, we  can  use our momentum space  results described 
in previous  sections  to arrive at  the following QCD predictions for both
coefficient functions\footnote{The details of the corresponding  Fourier transformation are 
spelled e.g. in  \cite{Chetyrkin:2010dx}.}.
\bea
&{}&\bar{\gamma}_{\theta;\unit}(r)=a_s^2 + a_s^3
\Biggl(
-\frac{1}{12} 
+\frac{11}{2}  \,l_{\mu X}
+
n_f 
\left[
-\frac{1}{18} 
-\frac{1}{3}  \,l_{\mu X}\,
\right]
\Biggr)
{+}  a_s^4\,\Biggl(
-\frac{49}{24} 
-\frac{495}{8}  \sbz \zeta_{3}
+\frac{397}{16}  \,l_{\mu X}\,
\nonumber\\
&{+}& 
\frac{363}{16}  \, l^2_{\mu X}\, 
+\,n_f 
\left[
-\frac{35}{144} 
+\frac{5}{4}  \sbz \zeta_{3}
-\frac{43}{12}  \,l_{\mu X}\,
-\frac{11}{4}  \, l^2_{\mu X}\, 
\right]
+n_f^2
\left[
-\frac{13}{216} 
+\frac{1}{36}  \,l_{\mu X}\,
+\frac{1}{12}  \, l^2_{\mu X}\, 
\right]
\Biggr)
\nonumber
\\
&{+}&  a_s^5\,\Biggl(
-\frac{255155}{1728} 
-\frac{2915}{16}  \sbz \zeta_{3}
+\frac{3465}{8}  \sbz \zeta_{5}
+\frac{20891}{192}  \,l_{\mu X}\,
-\frac{5445}{8}  \sbz \zeta_{3} \,l_{\mu X}\,
+\frac{1793}{8}  \, l^2_{\mu X}\, 
+\frac{1331}{16}  \, l^3_{\mu X}\, 
\nonumber\\
&{+}& \,n_f 
\left[
\frac{38741}{1728} 
-\frac{9}{16}  \sbz \zeta_{3}
-\frac{95}{36}  \sbz \zeta_{5}
-\frac{16685}{576}  \,l_{\mu X}\,
+55  \sbz \zeta_{3} \,l_{\mu X}\,
-\frac{4241}{96}  \, l^2_{\mu X}\, 
-\frac{121}{8}  \, l^3_{\mu X}\, 
\right]
\nonumber\\
&{+}& \, n_f^2
\left[
-\frac{361}{216} 
+\frac{125}{72}  \sbz \zeta_{3}
+\frac{491}{1728}  \,l_{\mu X}\,
-\frac{5}{6}  \sbz \zeta_{3} \,l_{\mu X}\,
+\frac{289}{144}  \, l^2_{\mu X}\, 
+\frac{11}{12}  \, l^3_{\mu X}\, 
\right]
\nonumber\\
&{+}& \, n_f^3
\left[
\frac{37}{1458} 
-\frac{1}{27}  \sbz \zeta_{3}
+\frac{13}{324}  \,l_{\mu X}\,
-\frac{1}{108}  \, l^2_{\mu X}\, 
-\frac{1}{54}  \, l^3_{\mu X}\, 
\right]
\Biggr)
+ {\mathcal O}(a_s^6)
\label{g0as5}
{},
\eea
\bea
&{}&\bar{\gamma}_{\theta;\theta}(r)= a_s^2\,\Biggl(
22- \frac{4}{3}\,n_f
\Biggr)
\nonumber\\
&+& a_s^3\,\Biggl(
{+}
\frac{788}{3} 
+121  \,l_{\mu X}\,
{+} \,n_f 
\left[
-\frac{304}{9} 
-\frac{44}{3}  \,l_{\mu X}\,
\right]
{+} \, n_f^2
\left[
\frac{8}{27} 
+\frac{4}{9}  \,l_{\mu X}\,
\right]
\Biggr)
+ {\mathcal O}(a_s^4) \label{g1as3}
{}.
\eea
Note that our {\em vacuum} calculations produce no information
about the coefficient function $\bar{\gamma}_{\theta;e+p}$ corresponding to the
{\em traceless} part of the energy-momentum tensor.

Numerically eqs.~(\ref{g0as5}) and \re{g1as3} read (we set $l_{\mu X}=0$)
\beq
\bar{\gamma}_{\theta;\unit}(r)\bbuildrel{=\!=\!=}_{n_f=0}^{} 
a_s^2 - 0.08333 \, \as^3  - 76.4189\, \as^4 + 82.4604  \as^5+ {\mathcal O}(a_s^6)
\label{}
{},
\eeq
\beq
\bar{\gamma}_{\theta;\unit}(r)\bbuildrel{=\!=\!=}_{n_f=3}^{} a_s^2 
- 0.25\, \as^3  -   73.1821\, \as^4 + 142.705  \as^5+ {\mathcal O}(a_s^6)
\label{}
{},
\eeq

\beq
\bar{\gamma}_{\theta;\theta}(r)  \bbuildrel{=\!=\!=}_{n_f=0}^{} 22\,\Biggl( a_s^2
+ 11.9394\, \as^3 \Biggr) + {\mathcal O}(a_s^4)
\label{}
{},
\eeq

\beq
\bar{\gamma}_{\theta;\theta}(r)  \bbuildrel{=\!=\!=}_{n_f=3}^{}
18 \, \Biggl(
 a_s^2
+  9.11111\, \as^3 
\Biggr)
+ {\mathcal O}(a_s^4)
\label{}
{}.
\eeq

\subsection{Shear stress correlator}

In \cite{Schroder:2011ht} the so-called shear stress correlator, defined as
\beq
G_\eta(X) = -16\,c_\eta^2  \,\langle   T[ T^{12}(X)\, T^{12}(0)] \rangle_c
\label{shear:X}
\eeq
with $X = (X_0,\vec{X}), \vec{X} = (0,0,X_3)$,
has been computed up to two-loops in high-temperature Yang-Mills theory.
Here $c_\eta$ is  an arbitrary constant (introduced for some reason that is not quite clear to us
in \cite{Schroder:2011ht}) which we put for simplicity equal to $\mathrm{i}/4$.
The calculation has been performed with the help of an ultraviolet expansion valid
in the limit of  small distances or large momenta; the result has been presented 
in the form of  an OPE. As the corresponding Wilson coefficients should  be $\bf T$-independent 
the results of  \cite{Schroder:2011ht} can be  checked and extended further with 
the help of our calculations.\footnote{The Wilson coefficients
in front of Lorentz non-invariant operators are for the moment not reachable with our projectors.
It would be interesting however to extend these methods in order to reach e.g. the coefficient
in front of $\langle T^{00} \rangle \sim e+p$. \ice{ with a similar approach.}
This is  possible in  {\em principle} with  the method of projectors as the  latter 
is  certainly not limited to the case of Lorenz-invariant operators in the rhs of an OPE. 
For example, in \cite{Larin:1996wd} the three-loop coeffcient functions of various tensor quark and gluon operators
of rank as large as 8  have been successfully  computed with the  help of  the method of  projectors.  

}

We start from momentum space. In the zero temperature limit the function 
\[
\tilde{G}_\eta(Q^2) = \int  \mathrm{d}^4 X \, e^{iQ X} \, G_\eta(X) 
\]
is related  to  contribution to energy-momentum tensor correlator  \re{TT}  proportional
to  the tensor structure $t_5^{\mu\nu;\rho\sigma}(q)$. This fact could be easily  checked 
by applying projector (2.5) of \cite{Schroder:2011ht} to the correlator 
$T^{\mu\nu;\rho\sigma}(q)$ expressed in terms of five  independent tensor structures 
displayed in \re{t1_t5}. The result reads
$-8\, Q^4\, ( 1- 7/2\, \eps +7/2\, \eps^2 - \eps^3)\,
\left(
C^5_{\unit}(Q^2) +  C^5_{\theta}(Q^2)\,\frac{\langle 0| \theta  |0 \rangle}{Q^4}  + \dots 
\right)
{}.
$

Thus, we will work  with the representation
\beq 
T^{\mu\nu;\rho\sigma}(q) \bbuildrel{=\!=\!=}_{q^2 \to -\infty}^{}  t_5^{\mu\nu;\rho\sigma}(q)
\Biggl(
C^5_{\unit}(Q^2) + C^5_{\theta}(Q^2)\,\frac{\langle 0| \theta  |0 \rangle}{Q^4}  + \dots 
\Biggr) + \mbox{ structures 1-4}
\label{C5:def}
\eeq

We first  concentrate on the  coefficient function  $C^5_{\theta}(Q^2)$ as the two-loop expression
for $C^5_{\unit}(Q^2)$ presented in \cite{Schroder:2011ht} is in  agreement to the previously known expression obtained in 
\cite{Pivovarov_tensorcurrents}. 
The result of \cite{Schroder:2011ht} for the second term in  eq.~\re{C5:def}
reads:
\bea
 C^5_{\theta}(Q^2) &=& -\frac{1}{3\beta_0\, a_s}\,
\Biggl(
1 - \frac{\beta_0\, \as}{4} \ln \zeta_{12} 
\Biggr)
{},
\label{C15:lattice:res}
\eea
where $\zeta_{12}$ is an unknown constant.
Note that the second term of the above expression is obtained not from a calculation but 
with the use of Renormalization Group considerations similar to those leading to
eq.~\re{RG:constraints}. Such a derivation assumes  that the coefficient function $C^5_{\theta}$ is finite
which is not obvious as the corresponding Feynman integrals have  logarithmic divergences
stemming from the region of small $x$ in    eq.~\re{TT}. Our direct calculation explicitly demonstrates
the presence of such divergences:
\bea
 C^{(5)}_{\theta}(Q^2) &= & \frac{1}{3\,\beta(\as)}\, \Biggl\{
1+ a_s\,\Biggl(\frac{41}{24} + \frac{7}{288}\, n_f\Biggr)
+ a_s^2\,\Biggl(\frac{117}{32} - \frac{457}{576} \, n_f + \frac{1}{576}\, n_f^2
\Biggr)
\nonumber
\\
&+& \frac{\as}{\eps}\Biggl(
\frac{11}{4} - \frac{1}{6} \, n_f
\Biggr)
+ \frac{\as^2}{\eps}\Biggl(
\frac{51}{8} - \frac{19}{24} \, n_f
\Biggr)
\Biggr\}
\label{C5}
{}.
\eea

It is important to note that  the contribution proportional to $C^5_{\theta}$ in \re{C5:def}
contains contact  terms {\em only}. This is in agreement with \re{C1Tfin}
due to an identity
\beq
C^{(T)}_1 - C^{(5)}_1 = \mbox{contact terms}
\label{C5=C1}
{},
\eeq
which, in turn,  follows from restriction \re{Tmumu:conservation} (recall that
$ C^{(5)}_{\theta}(Q^2) \equiv  C^{(5)}_1/{(-2\, \beta(\as))}$  as a 
consequence of \re{Tmumu_G2}).

In Euclidean position space eq.~\re{C5:def} can be presented as follows:
\beq 
\hat{T}^{\mu\nu;\rho\sigma}(X) \bbuildrel{=\!=\!=}_{r \to 0}^{}  
\Biggl(\delta^{\mu \rho} \delta^{\nu \sigma}+\delta^{\mu \sigma} \delta^{\nu \rho}\Biggr)
\Biggl\{
\tilde{C}^5_{\unit}(r)\, \unit + \tilde{C}^5_{\theta}(r)\,\langle 0| \theta  |0 \rangle  + \dots 
\Biggr\} + \mbox{ structures 1-4}
\label{xC5:def}
\eeq

Eq.~\re{C5=C1}, rewritten in terms of RG invariant quantities  assumes the form:
\beq
C^{(T)}_{1,RGI}  - 2\, \beta_0\, C^{(5)}_\theta = \mbox{contact terms}
\label{C5=C1:RG}
{}.
\eeq
By recalling that the contact terms do not contribute the function $G_\eta(x)$ for all $ x\not=0$ we 
conclude that eqs. \re{C5} and \re{CT:3l:log} contain all information to construct
the  first non-zero term ${\cal O}(\as^2)$ in the coefficient function  $\tilde{C}^5_{\theta}(x)$ with the result
\be
2\, \beta_0\tilde{C}^5_{\theta}(r) = 
\frac{\as^2}{\pi^2\, r^4} \Biggl(
\frac{107}{192} +\frac{17}{16} \, n_f - \frac{5}{48} \, n_f^2 + \frac{1}{5184}\, n_f^3
\Biggr)
\label{xC5:res}
\eeq

Finally, using the identity 
\beq
C^{(T)}_0 - C^{(5)}_0 = \mbox{contact terms}
\label{C5_0=C1_0}
{},
\eeq
and \re{adlerC0T} we arrive at the following result 

\bea
\tilde{C}^5_{\unit}(x)  &=& \frac{1}{\pi^4\, r^8}\Biggl\{ 
\frac{48}{5}+ 
 \frac{9}{5}\,n_f 
+ \as\, \Biggl(
-16
{+} 
\frac{7}{3}\,n_f 
\Biggr)
+ \as^2\, \Biggl(
\frac{711}{5} 
-\frac{1188}{5}  \sbz \zeta_{3}
-44  \,l_{\mu X}\,
\nonumber
\\
&{+}& \,n_f 
\left[
-\frac{259}{120} 
-\frac{27}{5}  \sbz \zeta_{3}
+\frac{109}{12}  \,l_{\mu X}\,
\right]
+ n_f^2
\left[
-\frac{41}{90} 
+\frac{6}{5}  \sbz \zeta_{3}
-\frac{7}{18}  \,l_{\mu X}\,
\right]
\Biggr)
\Biggr\}
\label{C0tx}
{}.
\eea

Numerical versions of  eqs. \re{xC5:res} and \re{C0tx} with $l_{\mu X} =0$ are presented below.
\bea
2\, \beta_0\, \tilde{C}^5_{\theta}(x)  \ \ &\bbuildrel{=\!=\!=}_{n_f=0}^{} &
\frac{\as^2}{\pi^2\, r^4}\Biggl\{ \frac{107}{192} = 0.557292\Biggr\}
{},
\\
2\, \beta_0\, \tilde{C}^5_{\theta}(x)  \ \ &\bbuildrel{=\!=\!=}_{n_f=3}^{}&
 \frac{\as^2}{\pi^2\, r^4}\Biggl\{\frac{45}{16}   = 2.81250
\Biggr\}
{},
\eea
\beq
\tilde{C}^5_{\unit}(x)  \ \ \bbuildrel{=\!=\!=}_{n_f=0}^{} \ \ \frac{48}{5}\frac{1}{\pi^4\, r^8}
\Biggl(
1  - 1.66667\, \as  -14.9384\, \as^2 
\Biggr)
{},
\eeq
\beq
\tilde{C}^5_{\unit}(x)  \ \ \bbuildrel{=\!=\!=}_{n_f=3}^{}\ \  \frac{15}{\pi^4\, r^8}
\Biggl(
1  -  0.6\, \as   -10.6983\, \as^2 
\Biggr)
{}.
\eeq

\section{Discussion and Conclusions}

We have presented higher order corrections to coefficient functions
$C_0$ and $C_1$ of the OPE of two energy-momentum tensors in massless
QCD as well as for the OPE of two scalar ``gluon condensate''
operators in massless QCD.  Our results extend the previously known
accuracy by one loop for the coefficient functions in front of the unit operator and by
two loops for the CF of the gluon condensate operator $O_1=-\f{1}{4}
G^{\mu \nu}G_{\mu \nu}$.

We have confirmed all previously available results and in some cases
extended them from purely Yang-Mills theory to QCD. Contrary to
previous assumptions, we have found that the coefficient functions  $C_1^{GG}$ as well as
$C^{(5)}_{\theta}(Q^2)$ are not completely finite with the standard
QCD renormalization.

We  thank  H. B. Meyer who has drawn  our attention to the importance of  
the energy-momentum tensor correlator.
Furthermore we would like to thank Y. Schr\"oder, A. Vuorinen, M. Laine and M. Jamin for useful comments.

We are grateful to J.~H.~K\"uhn for interesting discussions and support.

In conclusion we want to mention that all our calculations have been
performed on a SGI ALTIX 24-node IB-interconnected cluster of 8-cores
Xeon computers using the thread-based \cite{Tentyukov:2007mu} version  of FORM
\cite{Vermaseren:2000nd}.  The Feynman diagrams  have been drawn with the 
Latex package Axodraw \cite{Vermaseren:1994je}.

This work has been supported by the Deutsche Forschungsgemeinschaft in the
Sonderforschungsbereich/Transregio SFB/TR-9 ``Computational Particle
Physics''.

\begin{appendix}
 \section{Results for $C_0^{(r)}$ and $C_1^{(r)}$, $r=1\ldots 5$ and conversion to $C_0^{(S,T)}$ and $C_1^{(S,T)}$}
 Here we give our intermediate results for the coefficients $C_0^{(r)}$ and $C_1^{(r)}$ ($r=1\ldots 5$)
 appearing in the first line of eq.~(\ref{Ci2ten}), i.e. the coefficients for the tensor structures
 eq.~(\ref{t1_t5}) before subtraction of contact terms:
 \bea
 C_0^{(r)}&=C_0^{B\,(r)},\\
 C_1^{(r)}&=\f{1}{Z_G^2}C_1^{B\,(r)}.
 \eea
The conservation of the energy-momentum tensor in classical field theory translates to
\be
\partial_\mu\,T^{\mu\nu}=\text{local terms \cite{nielsen_Tmunu}.}
\ee
From this we get the relation
\be 
q_\mu C_{i}^{\mu\nu;\rho\sigma}(q)=\text{(local) contact terms}\quad (i=0,1)
\ee
which leads to the three restrictions serving as checks in our calculation:
\be \begin{split}
D_{i}^{(1)}(Q^2)&:=C_{i,1}(Q^2)+C_{i,2}(Q^2)+2\,C_{i,3}(Q^2)=\text{(local) contact terms},\\
D_{i}^{(2)}(Q^2)&:=C_{i,2}(Q^2)+C_{i,4}(Q^2)=\text{(local) contact terms},\\
D_{i}^{(3)}(Q^2)&:=C_{i,3}(Q^2)+C_{i,5}(Q^2)=\text{(local) contact terms}.\\
    \end{split} \label{Di123} \ee
Hence the subtraction of contact terms enables us to write the Wilson coefficients in terms of only two independent tensor
structures eq.~(\ref{tST}) which are related to the original five by the following equations:
\be \begin{split}
C_{i}^{(S)}(Q^2)&=-C_{i,2}(Q^2)-\f{2}{(D-1)}C_{i,3}(Q^2),\\
C_{i}^{(T)}(Q^2)&=-C_{i,3}(Q^2).
\end{split} \label{Ci25ten}\ee

 \subsection{\bf $C_0^{(r)}$, $r=1\ldots 5$}
These coefficients fulfill the relations eq.~(\ref{Di123}) even without local terms:
\be \begin{split}
D_{0}^{(1)}(Q^2)&:=C_{0,1}(Q^2)+C_{0,2}(Q^2)+2\,C_{0,3}(Q^2)=0,\\
D_{0}^{(2)}(Q^2)&:=C_{0,2}(Q^2)+C_{0,4}(Q^2)=0,\\
D_{0}^{(3)}(Q^2)&:=C_{0,3}(Q^2)+C_{0,5}(Q^2)=0.\\
    \end{split} \label{D0123} \ee
Hence it is enough to give $C_{0,4}$ and $C_{0,5}$ here:
\be \begin{split}
\lb16\pi^2\rb   C_{0,4} &=
       + \f{1}{\eps^2}   \left\{
          - \f{11}{1944} \as^2 \ca^2 \Ng
          + \f{109}{15552} \as^2 \Nf \ca \tr \Ng
          - \f{7}{3888} \as^2 \Nf^2 \tr^2 \Ng
          \right\}\\
      &\;\; + \f{1}{\eps}   \left\{
          - \f{1}{15} \Ng
          - \f{1}{30} \Nf \dR
          + \f{1}{54} \as \ca \Ng
          - \f{7}{432} \as \Nf \tr \Ng 
          - \f{35}{11664} \as^2 \ca^2 \Ng \right.\\ &\left. \qquad\quad\,
          + \f{1}{192} \as^2 \Nf \tr \cf \Ng
          - \f{809}{93312} \as^2 \Nf \ca \tr \Ng
          + \f{77}{23328} \as^2 \Nf^2 \tr^2 \Ng
          \right\}\\
      &\;\; - \f{47}{450} \Ng
          - \f{23}{225} \Nf \dR
          - \f{1}{15} l_{\sss{\mu q}} \Ng
          - \f{1}{30} l_{\sss{\mu q}} \Nf \dR\\ &\;\;
+\as\left\{
          + \f{187}{1620}  \ca \Ng
          - \f{1987}{12960}  \Nf \tr \Ng
          + \f{1}{27}  l_{\sss{\mu q}} \ca \Ng
          - \f{7}{216}  l_{\sss{\mu q}} \Nf \tr \Ng \right.\\ &\left.\qquad\quad\;
          + \f{1}{5} \zeta_{3}  \ca \Ng
          + \f{1}{10} \zeta_{3}  \Nf \tr \Ng \right\}\\ &\;\;
+\as^2\left\{
          + \f{160831}{1399680}  \ca^2 \Ng
          + \f{61}{480}  \Nf \tr \cf \Ng
          - \f{1733639}{2799360}  \Nf \ca \tr \Ng
          + \f{140909}{699840}  \Nf^2 \tr^2 \Ng \right.\\ &\left.\qquad\quad\;
          + \f{941}{9720}  l_{\sss{\mu q}} \ca^2 \Ng
          + \f{1}{64}  l_{\sss{\mu q}} \Nf \tr \cf \Ng
          - \f{15943}{77760}  l_{\sss{\mu q}} \Nf \ca \tr \Ng
          + \f{593}{9720}  l_{\sss{\mu q}} \Nf^2 \tr^2 \Ng \right.\\ &\left.\qquad\quad\;
          - \f{11}{648}  l_{\sss{\mu q}}^2 \ca^2 \Ng
          + \f{109}{5184}  l_{\sss{\mu q}}^2 \Nf \ca \tr \Ng
          - \f{7}{1296}  l_{\sss{\mu q}}^2 \Nf^2 \tr^2 \Ng \right.\\ &\left.\qquad\quad\;
          - \f{5}{12} \zeta_{5}  \ca^2 \Ng
          - \f{1}{4} \zeta_{5}  \Nf \tr \cf \Ng
          + \f{1}{24} \zeta_{5}  \Nf \ca \tr \Ng \right.\\ &\left.\qquad\quad\;
          + \f{563}{720} \zeta_{3}  \ca^2 \Ng
          + \f{37}{240} \zeta_{3}  \Nf \tr \cf \Ng
          - \f{29}{720} \zeta_{3}  \Nf \ca \tr \Ng
          - \f{19}{180} \zeta_{3}  \Nf^2 \tr^2 \Ng \right.\\ &\left.\qquad\quad\;
          + \f{11}{60} \zeta_{3}  l_{\sss{\mu q}} \ca^2 \Ng
          + \f{1}{40} \zeta_{3}  l_{\sss{\mu q}} \Nf \ca \tr \Ng
          - \f{1}{30} \zeta_{3}  l_{\sss{\mu q}} \Nf^2 \tr^2 \Ng \right\},
\end{split} \ee
\be \begin{split}
\lb16\pi^2\rb   C_{0,5} &=
       + \f{1}{\eps^2}   \left\{
          + \f{11}{1296} \as^2 \ca^2 \Ng
          - \f{109}{10368} \as^2 \Nf \ca \tr \Ng
          + \f{7}{2592} \as^2 \Nf^2 \tr^2 \Ng
          \right\}\\
      &\;\; + \f{1}{\eps}   \left\{
          + \f{1}{10} \Ng
          + \f{1}{20} \Nf \dR
          - \f{1}{36} \as \ca \Ng
          + \f{7}{288} \as \Nf \tr \Ng 
          - \f{1}{864} \as^2 \ca^2 \Ng \right.\\ &\left. \qquad\quad\,
          - \f{1}{128} \as^2 \Nf \tr \cf \Ng
          + \f{415}{20736} \as^2 \Nf \ca \tr \Ng
          - \f{35}{5184} \as^2 \Nf^2 \tr^2 \Ng
          \right\}\\
      &\;\; + \f{9}{100} \Ng
          + \f{3}{25} \Nf \dR
          + \f{1}{10} l_{\sss{\mu q}} \Ng
          + \f{1}{20} l_{\sss{\mu q}} \Nf \dR\\ &\;\;
+\as\left\{
          - \f{1}{540}  \ca \Ng
          + \f{1367}{8640}  \Nf \tr \Ng
          - \f{1}{18}  l_{\sss{\mu q}} \ca \Ng 
          + \f{7}{144}  l_{\sss{\mu q}} \Nf \tr \Ng \right.\\ &\left.\qquad\quad\;
          - \f{3}{10} \zeta_{3}  \ca \Ng
          - \f{3}{20} \zeta_{3}  \Nf \tr \Ng \right\}\\ &\;\;
+\as^2\left\{
          + \f{343429}{933120}  \ca^2 \Ng
          - \f{307}{1440}  \Nf \tr \cf \Ng
          + \f{983059}{1866240}  \Nf \ca \tr \Ng
          - \f{109129}{466560}  \Nf^2 \tr^2 \Ng \right.\\ &\left.\qquad\quad\;
          - \f{67}{12960}  l_{\sss{\mu q}} \ca^2 \Ng
          - \f{3}{128}  l_{\sss{\mu q}} \Nf \tr \cf \Ng
          + \f{10663}{51840}  l_{\sss{\mu q}} \Nf \ca \tr \Ng
          - \f{473}{6480}  l_{\sss{\mu q}} \Nf^2 \tr^2 \Ng \right.\\ &\left.\qquad\quad\;
          + \f{11}{432}  l_{\sss{\mu q}}^2 \ca^2 \Ng
          - \f{109}{3456}  l_{\sss{\mu q}}^2 \Nf \ca \tr \Ng
          + \f{7}{864}  l_{\sss{\mu q}}^2 \Nf^2 \tr^2 \Ng \right.\\ &\left.\qquad\quad\;
          + \f{5}{8} \zeta_{5}  \ca^2 \Ng
          + \f{3}{8} \zeta_{5}  \Nf \tr \cf \Ng
          - \f{1}{16} \zeta_{5}  \Nf \ca \tr \Ng \right.\\ &\left.\qquad\quad\;
          - \f{563}{480} \zeta_{3}  \ca^2 \Ng
          - \f{37}{160} \zeta_{3}  \Nf \tr \cf \Ng
          + \f{29}{480} \zeta_{3}  \Nf \ca \tr \Ng
          + \f{19}{120} \zeta_{3}  \Nf^2 \tr^2 \Ng \right.\\ &\left.\qquad\quad\;
          - \f{11}{40} \zeta_{3}  l_{\sss{\mu q}} \ca^2 \Ng
          - \f{3}{80} \zeta_{3}  l_{\sss{\mu q}} \Nf \ca \tr \Ng
          + \f{1}{20} \zeta_{3}  l_{\sss{\mu q}} \Nf^2 \tr^2 \Ng \right\}.
\end{split} \ee

 \subsection{\bf $C_1^{(r)}$, $r=1\ldots 5$}
Here we give the five coefficients
\be \begin{split}
   C_{1,1} =&\as\left\{
        \f{4}{9}   \ca
      - \f{7}{18}   \Nf \tr \right\}\\
            &+\as^2\left\{
        \f{3}{8}  \Nf \tr \cf
      + \f{1}{36} \Nf \ca \tr
      - \f{1}{18} \Nf^2 \tr^2  \right\},\\
\end{split}\ee 
\be \begin{split}
   C_{1,2} =&\as\left\{
       -   \ca
          + \f{1}{4}   \Nf \tr \right\} \\    
             &+\as^2\left\{
      -  \f{83}{216}  \ca^2
      + \f{25}{48} \Nf \tr \cf
      + \f{35}{216} \Nf \ca \tr 
      + \f{5}{108} \Nf^2 \tr^2  \right\},\\
\end{split}\ee 
\be \begin{split}
   C_{1,3} =&\as\left\{
        \f{5}{18}   \ca
          + \f{5}{72}   \Nf \tr \right\}  \\ 
               &+\as^2\left\{
        \f{83}{432}  \ca^2
      - \f{43}{96} \Nf \tr \cf
      - \f{41}{432} \Nf \ca \tr 
      + \f{1}{216} \Nf^2 \tr^2  \right\},\\
\end{split}\ee 
\be \begin{split}
   C_{1,4} =
       &+\f{1}{3}\\
       & \f{\as}{\eps}   \left\{
           \f{11}{36}   \ca
          - \f{1}{9}   \Nf \tr
          \right\}\\
       &+ \f{1}{3} +\as\left\{
           \f{161}{216}   \ca
          - \f{17}{108}  \Nf \tr\right\}   \\ 
       & +\f{\as^2}{\eps}   \left\{
           \f{17}{72}   \ca^2
         - \f{1}{12}   \Nf \tr \cf
         - \f{5}{36}   \Nf \ca \tr
          \right\}\\
       &+\as^2\left\{
        \f{3}{16}  \ca^2
      - \f{65}{144} \Nf \tr \cf
      - \f{5}{108} \Nf \ca \tr 
      - \f{5}{108} \Nf^2 \tr^2  \right\},\\
\end{split}\ee 
\be \begin{split}
   C_{1,5} =
       &-\f{2}{3}\\
       & \f{\as}{\eps}   \left\{
          - \f{11}{18}  \as \ca
          + \f{2}{9}  \as \Nf \tr
          \right\}\\
       &- \f{2}{3} +\as\left\{
          - \f{41}{108}   \ca
          - \f{7}{216}  \Nf \tr\right\}\\
       & +\f{\as^2}{\eps}   \left\{
         - \f{17}{36}   \ca^2
         + \f{1}{6}   \Nf \tr \cf
         + \f{5}{18}   \Nf \ca \tr
          \right\}\\
       &+\as^2\left\{
      -  \f{13}{48}  \ca^2
      + \f{137}{288} \Nf \tr \cf
      + \f{61}{432} \Nf \ca \tr 
      - \f{1}{216} \Nf^2 \tr^2  \right\},\\
\end{split}\ee  
which fulfill the relations eq.~(\ref{Di123}):
\be \begin{split}
D_{1}^{(1)}  =&\, 0\,,\\
D_{1}^{(2)}   =&\text{local } \neq 0,\\
 D_{1}^{(3)}  =&\text{local } \neq 0.
\end{split} \label{D1123}\ee 
This is an important check as for the coefficient $C_1^{\mu\nu;\rho\sigma}$ only
counterterms of the form $t_{4}^{\mu\nu;\rho\sigma} [O_1]$ and $t_{5}^{\mu\nu;\rho\sigma} [O_1]$ are possible.
Counterterms proportional to the other tensor structures would not be local.
Hence $D_{1}^{(1)} =0$ is necessary.
\end{appendix}

\bibliographystyle{JHEP}

\bibliography{Literatur_v2}

\end{document}